\documentclass[12pt,a4paper]{article}
\usepackage{amsmath,amssymb,amsthm} 
\numberwithin{equation}{section}

\theoremstyle{remark}
\newtheorem{remark}{Remark}
\newcommand{\beq}{\begin{equation}}
\newcommand{\eeq}{\end{equation}}
\newcommand{\beqnn}{\begin{equation*}}
\newcommand{\eeqnn}{\end{equation*}}
\newcommand{\rd}{\partial}

\newcommand{\diag}{\operatorname{diag}}
\newcommand{\tp}[1]{\,{\vphantom{#1}}^\mathrm{t}\!\,#1}
\newcommand{\CC}{\mathbb{C}}
\newcommand{\PP}{\mathbb{P}}

\newcommand{\ZZ}{\mathbb{Z}}
\newcommand{\calH}{\mathcal{H}}
\newcommand{\calN}{\mathcal{N}}
\newcommand{\calP}{\mathcal{P}}
\newcommand{\bsa}{\boldsymbol{a}}
\newcommand{\bst}{\boldsymbol{t}}
\newcommand{\bsx}{\boldsymbol{x}}

\newcommand{\bsT}{\boldsymbol{T}}
\newcommand{\bsV}{\boldsymbol{V}}
\newcommand{\bsGamma}{\boldsymbol{\Gamma}}
\newcommand{\frakL}{\mathfrak{L}}
\newcommand{\gl}{\mathrm{gl}}
\newcommand{\GL}{\mathrm{GL}}
\newcommand{\bszero}{\boldsymbol{0}}

\begin{document}

\title{Toda hierarchies and their applications}
\author{Kanehisa Takasaki\thanks{E-mail: takasaki@math.kindai.ac.jp}\\
{\normalsize Department of Mathematics, Kindai University}\\ 
{\normalsize 3-4-1 Kowakae, Higashi-Osaka, Osaka 577-8502, Japan}}
\date{}
\maketitle 

\begin{abstract}
The 2D Toda hierarchy occupies a central position in the family 
of integrable hierarchies of the Toda type. The 1D Toda hierarchy 
and the Ablowitz-Ladik (aka relativistic Toda) hierarchy 
can be derived from the 2D Toda hierarchy as reductions.  
These integrable hierarchies have been applied to various problems 
of mathematics and mathematical physics since 1990s.  
A recent example is a series of studies on models 
of statistical mechanics called the melting crystal model. 
This research has revealed that the aforementioned two reductions 
of the 2D Toda hierarchy underlie two different melting crystal models.  
Technical clues are a fermionic realization of the quantum torus algebra, 
special algebraic relations therein called shift symmetries, and 
a matrix factorization problem.  The two melting crystal models thus 
exhibit remarkable similarity with the Hermitian and unitary matrix models 
for which the two reductions of the 2D Toda hierarchy play the role 
of fundamental integrable structures. 
\end{abstract}

\begin{flushleft}
\end{flushleft}

\newpage

\section{Introduction}

In 1967, Morikazu Toda introduced a one-dimensional lattice 
mechanical system with exponential interactions 
nowadays called the Toda lattice \cite{Toda67a}. 
Though designed to have a periodic solution written 
in terms of elliptic functions \cite{Toda67b}, 
this nonlinear lattice was soon shown to have a solution 
with colliding solitons.  This suggested remarkable similarity 
with the KdV equation, hence integrability.  

Integrability of the Toda lattice was established 
by the middle of 1970s after the construction 
of exact $N$-soliton solutions \cite{Hirota73}, 
first integrals in involution \cite{Henon74,Flaschka74a}, 
Lax pairs for the inverse scattering method \cite{Flaschka74b,Manakov74} 
and finite-band integration of the periodic problem \cite{KvM75,DT76}. 
These results were extended to a system of two-dimensional 
relativistic fields with exponential interactions 
among the components of the fields.  
By the end of 1970s, this 2D Toda field equation was proved 
to be integrable by the Lie group theory \cite{LS79}, 
the inverse scattering method \cite{Mikhailov79,Mikhailov81} 
and the bilinearization method \cite{Hirota81}.  
In the beginning of 1980s, a fully 3D discretization 
was proposed in a bilinear form along with $N$-soliton solutions 
\cite{Hirota81}.  A description of more general solutions 
of this discrete system was soon presented in the language 
of a 2D complex free fermion system \cite{Miwa82}.  

The Toda hierarchies \cite{UT84} were developed 
as a Toda version of the KP hierarchy \cite{SS83,SW85} 
and its various relatives \cite{JM83}. 
One of its prototypes is an unpublished result of van Moerbeke 
that is quoted in the work of Adler \cite{Adler79}.  
This result explains how to construct an integrable hierarchy 
of Lax equations for a difference operator $\frakL$.  
In particular, the integrable hierarchy for the Jacobi operator 
\beqnn
  \frakL = e^{\rd_s} + b + ce^{-\rd_s}, 
\eeqnn
referred to as the {\it 1D Toda hierarchy\/}, contains 
the equation of motion of the Toda lattice 
as the lowest member of the Lax equations 
with time variables $\bst = (t_1,t_2,\ldots)$. 
In view of the construction of the KP hierarchy 
with a pseudo-differential Lax operator, 
it is natural to extend this construction to 
a ``pseudo-difference operator'' of the form 
\beqnn
  L = e^{\rd_s} + u_0 + u_1e^{-\rd_s} + \cdots. 
\eeqnn
This extension, however, is not enough to accommodate 
the 2D Toda field equation.  To this end, 
another Lax operator of the form 
\beqnn
  \bar{L}^{-1} = \bar{u}_0e^{-\rd_s} + \bar{u}_1 
    + \bar{u}_2e^{\rd_s} + \cdots 
\eeqnn
has to be introduced along with another set $\bar{\bst} 
= (\bar{t}_1,\bar{t}_2,\ldots)$ of time variables. 
The {\it 2D Toda hierarchy} consists of Lax equations 
for these two Lax operators $L,\bar{L}$ with respect to 
the two sets $\bst,\bar{\bst}$ of time variables.  
The whole system of these Lax equations turns out 
to be equivalent to a system of Zakharov-Shabat equations 
for difference operators.  The 2D Toda field equation 
is contained therein as the lowest member.  

The 2D Toda hierarchy can be reformulated 
as a system of bilinear equations of the Hirota form 
for a single tau function $\tau(s,\bst,\bar{\bst})$ 
(in which the lattice coordinate $s$ is treated 
on an equal footing with the other independent variables). 
These bilinear equations can be cast into (and derived from) 
a generating functional form.  One can thereby deduce \cite{UT84} 
that $\tau(s,\bst,\bar{\bst})$, up to a sign factor, coincides 
with the tau function of the two-component KP hierarchy
with charge $(s,-s)$ \cite{DJKM81}.  
This leads to a fermionic formula of $\tau(s,\bst,\bar{\bst})$.
Actually, $\tau(s,\bst,\bar{\bst})$ has another fermionic formula 
\cite{Takebe91a,Takebe91b,AZ12} that is directly related 
to a matrix factorization problem for solving 
the 2D Toda hierarchy in the Lax formalism \cite{Takasaki84}. 
This fermionic formula is a very powerful tool for studying 
various special solutions of the 2D Toda hierarchy 
including those that we consider in this paper. 

Many applications of the 2D Toda hierarchy 
and its relatives have been found in mathematics 
and mathematical physics.  In the 1990s, 
the 1D and 2D Toda hierarchies were applied to 2D gravity 
\cite{GMMMO91,Martinec91,AGGJ91,KMMM92} 
and $c = 1$ string theory 
\cite{DMP93,HOP94,EK94,Takasaki95,NTT95,Takasaki96} 
as well as mathematical aspects of random matrices 
and orthogonal polynomials \cite{AvM95,KMZ96,AvM99}. 
This is also the place where the Ablowitz-Ladik hierarchy 
\cite{AL75} (aka the relativistic Toda hierarchy 
\cite{Ruijsenaars90}) plays a role.  
These studies also revealed new features 
of the 2D Toda hierarchy itself such as 
the Orlov-Schulman operators, additional symmetries 
and dispersionless analogues \cite{TT93,TT95}. 
Researches on the dispersionless 2D Toda hierarchy 
revived later on when a relation to interface dynamics 
and complex analysis was pointed out \cite{WZ00,MWWZ00}. 

Sources of new researches were discovered in the early 2000s 
in enumerative geometry of $\CC\PP^1$ and $\CC^2$ 
\cite{Pandharipande02,Okounkov00,Getzler01,OP02a,OP02b,DZ04,LQW03,QW04,Milanov05} 
and 4D $\calN=2$ supersymmetric gauge theories 
\cite{LMN03,Nekrasov02,NO03,MN06}.  For example, 
a generating function of the double Hurwitz numbers 
of $\CC\PP^1$ was shown to be a tau function 
of the 2D Toda hierarchy \cite{Okounkov00}.  
This tau function falls into a class of special tau functions 
called ``hypergeometric tau functions'' that was introduced 
around 2000 in a quite different context \cite{OS00,OS01a,OS01b}.  
Intersection numbers of the Hilbert scheme of points on $\CC^2$, 
too, give a hypergeometric tau function \cite{LQW03,QW04}. 
On the other hand, Gromov-Witten invariants of $\CC\PP^1$ 
yield a different kind of tau functions \cite{OP02a,OP02b}. 

This paper reviews our work in the last ten years 
on integrable structures of the melting crystal models 
\cite{NT07,NT08,NT11,Takasaki13,Takasaki14}.  
We focus on two typical cases among these models 
of statistical mechanics. The first case is a statistical model 
of random 3D Young diagrams \cite{ORV03}
(hence referred to as a ``crystal model'').  
Its partition function may be also thought of 
as the simplest instanton partition function 
of 5D supersymmetric gauge theories \cite{MNTT04}. 
The second case is a slight modification of the first case, 
and related to enumerative geometry \cite{BP08} 
and topological string theory \cite{CGMPS06} 
of a Calabi-Yau threefold called the ``resolved conifold''. 
Our work have proved that the 1D Toda hierarchy 
and the Ablowitz-Ladik hierarchy underlie 
these two melting crystal models.  Let us mention 
that such a relation between the resolved conifold 
and the Ablowitz-Ladik hierarchy was pointed out 
first by Brini \cite{Brini10}.  It is remarkable 
that these two integrable hierarchies, both of which are 
reductions of the 2D Toda hierarchy \cite{UT84,KMZ96,BCR11}, 
are also known to be the integrable structures 
of two typical random matrix models, namely 
the Hermitian and unitary random matrix models 
\cite{GMMMO91,Martinec91,AGGJ91,KMMM92,AvM95,KMZ96,AvM99}. 
Technical clues of our work are the quantum torus algebra 
in the fermionic formalism, special algebraic relations 
in this algebra referred to as ``shift symmetries'', 
and the matrix factorization problem in the Lax formalism. 

This paper is organized as follows.  
Section 2 is a review of the 2D Toda hierarchy formulated 
in the Lax and bilinear forms.  Fundamental building blocks 
of the 2D Toda hierarchy such as the Lax operators, 
the dressing operators, the wave functions and the tau function 
are introduced along with various equations.  
The 1D Toda and Ablowitz-Ladik hierarchies are shown 
to be reductions of the 2D Toda hierarchy.  
The matrix factorization problem is also commented.  
Section 3 is a review of the fermionic formalism 
of the 2D Toda hierarchy.  
The fermionic formula of the tau function and its relation 
to the matrix factorization problem are explained.  
Relevant combinatorial notions such as partitions, 
Young diagrams and the Schur functions are also introduced here.  
The fermionic formula is illustrated 
for hypergeometric tau functions, in particular, 
the generating function of the double Hurwitz numbers.  
Sections 4 and 5 are devoted to the melting crystal models.  
In Section 4, the two melting crystal models are introduced. 
The partition functions are defined as sums of 
the Boltzmann weights over the set of all partitions.  
Fermionic expressions of these partition functions are also derived.  
In Section 5, integrable structures of the two melting crystal models 
are identified.  The quantum torus algebra and its shift symmetries 
are reviewed.  With the aid of these algebraic tools, 
the partition functions are converted to tau functions 
of the 2D Toda hierarchy.  The first model thus turns out 
to be related to the 1D Toda hierarchy.  The second model 
is further examined in the Lax formalism, and shown 
to be related to the Ablowitz-Ladik hierarchy.  
Section 6 concludes these reviews.

\section{2D Toda hierarchy}

\subsection{Difference operators and infinite matrices}

The Lax formalism of the 2D Toda hierarchy is formulated 
by difference operators in the lattice coordinate $s$ \cite{UT84}.  
These operators are linear combinations of the shift operators $e^{n\rd_s}$ 
(symbolically expressed as the exponential of $\rd_s = \rd/\rd s$) 
that act on functions of $s$ as $e^{n\rd_s}f(s) = f(s+n)$.  
A genuine difference operators is a finite linear combination 
\beqnn
  A = \sum_{n=M}^N a_n(s)e^{n\rd_s} 
  \quad \mbox{(operator of $[M,N]$-type)}
\eeqnn
of the shift operators.  To formulate the 2D Toda hierarchy, 
we further use semi-infinite linear combinations of the form 
\beqnn
  A = \sum_{n=-\infty}^N a_n(s)e^{n\rd_s}
  \quad \mbox{(operator of $(-\infty,N]$ type}) 
\eeqnn
and 
\beqnn
  A = \sum_{n=M}^\infty a_n(s)e^{n\rd_s}
  \quad \mbox{(operator of $[M,\infty)$ type)}. 
\eeqnn
These ``pseudo-difference operators'' are analogues 
of pseudo-differential operators in the Lax formalism 
of the KP hierarchy \cite{SS83,SW85}. 
Let $(\quad)_{\ge 0}$ and $(\quad)_{<0}$ denote the projection 
\beqnn
  (A)_{\ge 0} = \sum_{n\ge 0}a_n(s)e^{n\rd_s}, \quad 
  (A)_{<0} = \sum_{n<0}a_n(s)e^{n\rd_s}. 
\eeqnn
to the $[0,\infty)$ and $(-\infty,-1]$ parts. 

These difference operators are also represented 
by $\ZZ\times\ZZ$ matrices.  The shift operators $e^{n\rd_s}$ 
correspond to the shift matrices 
\beqnn
  \Lambda^n = (\delta_{i,j-n})_{i,j\in\ZZ}. 
\eeqnn
The multiplication operators $a(s)$ are represented 
by the diagonal matrices 
\beqnn
  \diag(a(s)) = (a(i)\delta_{ij})_{i,j\in\ZZ}. 
\eeqnn
Thus a general difference operator of the form 
\beqnn
  A = A(s,e^{\rd_s}) = \sum_{n\in\ZZ} a_n(s)e^{n\rd_s}
\eeqnn
is represented by the infinite matrix 
\beqnn
  A(\Delta,\Lambda) 
  = \sum_{n\in\ZZ}\diag(a_n(s))\Lambda^n 
  = \sum_{n\in\ZZ} (a_n(i)\delta_{i,j-n})_{i,j\in\ZZ}. 
\eeqnn

The shift operator $e^{\rd_s}$ and the multiplication operator $s$ 
satisfy the twisted canonical commutation relation 
\beq
  [e^{\rd_s},s] = e^{\rd_s}. 
  \label{tCCR}
\eeq
This commutation relation can be translated 
to the language of matrices as 
\beq
  [\Lambda,\Delta] = \Lambda, 
\eeq
where $\Delta$ denotes the the diagonal matrix 
\beqnn
  \Delta = \diag(s) = (i\delta_{ij})_{i,j\in\ZZ} 
\eeqnn
that represents the multiplication operator $s$.

\subsection{Lax and Zakharov-Shabat equations}

The Lax formalism of the 2D Toda hierarchy uses 
two Lax operators $L,\bar{L}$ 
\footnote{In the earliest work \cite{UT84}, 
these Lax operators were denoted by $L,M$. 
These notations have been changed to $L,\bar{L}$ 
so as to use $M$ for the Orlov-Schulman operators. 
Also note that the bar $\;\bar{}\;$ of $\bar{L}$, $\bar{t}_k$, 
$\bar{u}_n$, etc. does not mean complex conjugation. }
of type $(-\infty,1]$ and $[1,\infty)$.   
From the point of view of symmetry, 
it is better to consider $L$ and $\bar{L}^{-1}$ 
rather than $L$ and $\bar{L}$.  These operators 
admit freedom of gauge transformations 
$L \to e^{-f}\cdot L\cdot e^f$, 
$\bar{L} \to e^{-f}\cdot\bar{L}\cdot e^f$. 
We mostly use the gauge in which the leading coefficient 
of $L$ is equal to $1$: 
\beqnn
\begin{aligned}
  L &= e^{\rd_s} + \sum_{n=1}^\infty u_ne^{(1-n)\rd_s},\\
  \bar{L}^{-1} &= \bar{u}_0e^{-\rd_s} 
     + \sum_{n=1}^\infty \bar{u}_n e^{(n-1)\rd_s}. 
\end{aligned}
\eeqnn
The coefficients $u_n$ and $\bar{u}_n$ are functions 
$u_n(s,\bst,\bar{\bst})$ and $\bar{u}_n(s,\bst,\bar{\bst})$ 
of $s$ and the time variables $\bst,\bar{\bst}$. 
To simplify notations, however, we shall frequently suppress 
$\bst$ and $\bar{\bst}$ as $u_n = u_n(s)$ and $\bar{u}_n = \bar{u}_n(s)$. 

$L$ and $\bar{L}$ satisfy the Lax equations 
\beq
\begin{gathered}
  \frac{\rd L}{\rd t_n} = [B_n,L], \quad 
  \frac{\rd L}{\rd\bar{t}_n} = [\bar{B}_n,L], \\
  \frac{\rd\bar{L}}{\rd t_n} = [B_n,\bar{L}],\quad
  \frac{\rd\bar{L}}{\rd\bar{t}_n} = [\bar{B}_n,\bar{L}], 
\end{gathered}
\label{Laxeq}
\eeq
where $B_n$ and $\bar{B}_n$ are defined as 
\beqnn
  B_n = (L^n)_{\ge 0}, \quad 
  \bar{B}_n = (\bar{L}^{-n})_{<0}. 
\eeqnn
$B_n$ and $\bar{B}_n$, in turn, satisfy 
the Zakharov-Shabat equations
\beq
\begin{gathered}
  \frac{\rd B_n}{\rd t_m} - \frac{\rd B_m}{\rd t_n} + [B_m,B_n] = 0, \\
  \frac{\rd\bar{B}_n}{\rd\bar{t}_m} - \frac{\rd\bar{B}_m}{\rd\bar{t}_n} 
    + [\bar{B}_m,\bar{B}_n] = 0, \\
  \frac{\rd\bar{B}_n}{\rd t_m} - \frac{\rd B_m}{\rd\bar{t}_n} 
    + [B_m,\bar{B}_n] = 0. 
\end{gathered}
\label{ZSeq}
\eeq
Actually, the Lax equations and the Zakharov-Shabat equations 
are equivalent \cite{UT84}. 

Since 
\beqnn
  B_1 = e^{\rd_s} + u_1,\quad \bar{B}_1 = \bar{u}_0e^{-\rd_s}, 
\eeqnn
the lowest ($m = n = 1$) member of the third set of 
the Zakharov-Shabat equation reduces to the equations 
\beqnn
\begin{gathered}
  \frac{\rd u_1(s)}{\rd\bar{t}_1} + \bar{u}_0(s+1) - \bar{u}_0(s) = 0,\\
  - \frac{\rd\bar{u}(s)_0}{\rd t_1} + \bar{u}_0(s)(u_1(s) - u_1(s-1)) = 0.
\end{gathered}
\eeqnn
Upon parametrizing $u_1$ and $\bar{u}_0$ with new dependent 
variable $\phi(s) = \phi(s,\bst,\bar{\bst})$ as  
\beqnn
  u_1(s) = \frac{\rd\phi(s)}{\rd t_1},\quad 
  \bar{u}_0(s) = e^{\phi(s) - \phi(s-1)}, 
\eeqnn
these equations yields the 2D Toda field equation 
\beq
  \frac{\rd^2\phi(s)}{\rd t_1\rd\bar{t}_1} 
  + e^{\phi(s+1) - \phi(s)} - e^{\phi(s) - \phi(s-1)} = 0. 
  \label{2DTodaeq}
\eeq

\subsection{Dressing operators and wave functions}

The Lax operators $L,\bar{L}$ can be converted 
to the undressed form $e^{\rd_s}$ as
\beq
  L = We^{\rd_s}W^{-1},\quad 
  \bar{L} = \bar{W}e^{\rd_s}\bar{W}^{-1}
  \label{LLbar-WWbar}
\eeq
by dressing operators of the form 
\beqnn
  W = 1 + \sum_{n=1}^\infty w_ne^{-n\rd_s},\quad 
  \bar{W} = \sum_{n=0}^\infty\bar{w}_ne^{n\rd_s},\quad 
  \bar{w}_0 \not= 0.
\eeqnn
One can further choose $W,\bar{W}$ to satisfy 
the Sato equations
\beq
\begin{gathered}
  \frac{\rd W}{\rd t_k} = B_kW - We^{k\rd_s},\quad 
  \frac{\rd W}{\rd\bar{t}_k} = \bar{B}_kW,\\
  \frac{\rd\bar{W}}{\rd t_k} = B_k\bar{W},\quad 
  \frac{\rd\bar{W}}{\rd\bar{t}_k} = \bar{B}_k\bar{W} - We^{-k\rd_s}. 
\end{gathered}
\label{Satoeq}
\eeq
 
Upon substituting the expression 
\beqnn
  B_k = \left(We^{k\rd_s}W^{-1}\right)_{\ge 0},\quad 
  \bar{B}_k = \left(\bar{W}e^{-k\rd_s}\bar{W}^{-1}\right)^{-1}
\eeqnn
for $B_k$'s and $\bar{B}_k$'s, the Sato equations (\ref{Satoeq}) 
turn into the closed system of evolution equations 
\beq
\begin{gathered}
  \frac{\rd W}{\rd t_k} = - \left(We^{k\rd_s}W^{-1}\right)_{<0}W,\quad 
  \frac{\rd W}{\rd\bar{t}_k} 
    = \left(\bar{W}e^{-k\rd_s}\bar{W}\right)_{<0}W,\\
  \frac{\rd\bar{W}}{\rd t_k} 
    = \left(We^{k\rd_s}W^{-1}\right)_{\geq 0}\bar{W},\quad 
  \frac{\rd\bar{W}}{\rd\bar{t}_k} 
    = - \left(\bar{W}e^{-k\rd_s}\bar{W}^{-1}\right)_{\geq 0}\bar{W} 
\end{gathered}
\label{Satoeq2}
\eeq
for $W$ and $\bar{W}$.  These equations and may be thought of 
as yet another formulation of the 2D Toda hierarchy, 
from which the Lax equations (\ref{Laxeq}) can be recovered 
through the relation (\ref{LLbar-WWbar}).  

The dressing operators can be used to define 
the wave functions 
\beq
  \Psi = \left(1 + \sum_{k=1}^\infty w_kz^{-k}\right)z^se^{\xi(\bst,z)},\quad 
  \bar{\Psi} 
    = \left(\sum_{k=0}^\infty\bar{w}_kz^k\right)z^se^{\xi(\bar{\bst},z^{-1})},
\eeq
where 
\beqnn
  \xi(\bst,z) = \sum_{k=1}^\infty t_kz^k,\quad 
  \xi(\bar{\bst},z^{-1}) = \sum_{k=1}^\infty\bar{t}_kz^{-k}. 
\eeqnn
The wave functions satisfy the auxiliary linear equations 
\beq
  L\Psi = z\Psi,\quad \bar{L}\bar{\Psi} = z\bar{\Psi}
  \label{LLbar-Lineq}
\eeq
and 
\beq
\begin{gathered}
  \frac{\rd\Psi}{\rd t_k} = B_k\Psi,\quad 
  \frac{\rd\Psi}{\rd\bar{t}_k} = \bar{B}_k\Psi,\\
  \frac{\rd\bar{\Psi}}{\rd t_k} = B_k\bar{\Psi},\quad 
  \frac{\rd\bar{\Psi}}{\rd\bar{t}_k} = \bar{B}_k\bar{\Psi}. 
\end{gathered}
\eeq

\subsection{Tau functions and bilinear equations}

The tau function $\tau = \tau(s,\bst,\bar{\bst})$ 
of the 2D Toda hierarchy is related to the wave functions as
\footnote{These relations differ from those commonly used 
in the literature.  We have replaced $\tau(s,\bst,\bar{\bst})$ 
therein by $\tau(s-1,\bst,\bar{\bst})$ so as to consistent 
with the convention of our fermionic formalism.}
\beq
\begin{gathered}
  \Psi(s,\bst,\bar{\bst},z) 
  = \frac{\tau(s-1,\bst-[z^{-1}],\bar{\bst})}{\tau(s-1,\bst,\bar{\bst})}
    z^se^{\xi(\bst,z)},\\
  \bar{\Psi}(s-1,\bst,\bar{\bst},z) 
  = \frac{\tau(s,\bst,\bar{\bst}-[z])}{\tau(s-1,\bst,\bar{\bst})}
    z^se^{\xi(\bar{\bst},z^{-1})},
\end{gathered}
\label{Psi-tau-rel}
\eeq
where 
\beqnn
  [z] = \left(z,z^2/2,\cdots,z^k/k,\cdots\right). 
\eeqnn
Given the pair $\Psi,\bar{\Psi}$ of wave functions, 
one can define the tau function as a kind of potential 
that satisfy these relations.  

The tau function satisfies an infinite number of Hirota equations.  
The first three members of these Hirota equations read 
\beq
\begin{gathered}
 D_1\bar{D}_1\tau(s,\bst,\bar{\bst})\cdot\tau(s,\bst,\bar{\bst}) 
  + 2\tau(s+1,\bst,\bar{\bst})\tau(s-1,\bst,\bar{\bst}) = 0, \\
 (D_2 + D_1^2)
  \tau(s+1,\bst,\bar{\bst})\cdot\tau(s,\bst,\bar{\bst}) = 0, \\
 (\bar{D}_2 + \bar{D}_1^2)
  \tau(s,\bst,\bar{\bst})\cdot\tau(s+1,\bst,\bar{\bst}) = 0,
\end{gathered}
\label{Toda-Hirotaeq}
\eeq
where we have used Hirota's notation 
\begin{multline*}
  P(D_1,D_2,\ldots,\bar{D}_1,\bar{D}_2,\ldots)
  f(\bst,\bar{\bst})\cdot g(\bst,\bar{\bst})\\
  = \left.
    P(\rd'_1 - \rd_1,\, \rd'_2 - \rd_2,\,\ldots,\,
      \bar{\rd}'_1 - \bar{\rd}_1,\bar{\rd}'_2 - \bar{\rd}_2,\,\ldots) 
    f(\bst',\bar{\bst}')g(\bst,\bar{\bst})\right|_{\bst'=\bst}, 
\end{multline*}
where $\rd_k,\rd'_k,\bar{\rd}_k,\bar{\rd}'_k$ denote 
the derivatives $\rd_k = \rd/\rd t_k$, $\rd'_k = \rd/\rd t'_k$, 
$\bar{\rd}_k = \rd/\rd\bar{t}_k$, $\bar{\rd}'_k = \rd/\rd\bar{t}'_k$. 
The first equation of (\ref{Toda-Hirotaeq}) amounts 
to the 2D Toda field equation (\ref{2DTodaeq}). 
The infinite system of Hirota equations can be encoded 
to (and decoded from) the single bilinear equation 
\begin{multline}
  \oint\frac{dz}{2\pi i}z^{s'-s}e^{\xi(\bst'-\bst,z)} 
    \tau(s',\bst'-[z^{-1}],\bar{\bst}') 
    \tau(s,\bst+[z^{-1}],\bar{\bst}) \\
  = \oint\frac{dz}{2\pi i}z^{s'-s}e^{\xi(\bar{\bst}'-\bar{\bst},z^{-1})} 
      \tau(s'+1,\bst',\bar{\bst}'-[z]) 
      \tau(s-1,\bst,\bar{\bst}+[z]), 
\label{Toda-bilin-tau}
\end{multline}
where the symbol $\oint$ means extracting the ``residue'' 
of a (formal) Laurent series: 
\beqnn
  \oint\sum_{n=-\infty}^\infty\frac{dz}{2\pi i}a_nz^n = a_{-1}.
\eeqnn
Analytically, this symbol on the left side of the equation 
is understood to be the contour integral 
along a sufficiently large circle $|z| = R$, and 
that of the right side is the contour integral 
along a sufficiently small circle $|z| = R^{-1}$. 

Various bilinear equations for the tau function 
can be derived from (\ref{Toda-bilin-tau}) by specialization 
of $\bst',\bar{\bst}'$ and $s'$.  The Hirota equations 
(\ref{Toda-Hirotaeq}) are obtained by Taylor expansion 
of (\ref{Toda-bilin-tau}) at $\bst' = \bst$ 
and $\bar{\bst}' = \bar{\bst}$ to low orders 
upon letting $s' = s, s\pm 1$.  
More systematic derivation of Hirota equations 
uses the polynomials $S_n(\bst)$, $n = 0,1,\ldots$, 
defined by the generating function
\beq
  \sum_{n=0}^\infty S_n(\bst)z^n 
  = \exp\left(\sum_{k=1}^\infty t_kz^k\right). 
  \label{S_n}
\eeq
These polynomials are building blocks the Schur functions as well 
(we refer to Macdonald's book \cite{Mac-book} for the notions 
of the Schur functions, partitions and Young diagrams). 
Thus a complete set of Hirota equations can be obtained 
in the generating functional form
\begin{multline}
  \sum_{n=0}^\infty 
    S_n(-2\bsa)S_{n+s'-s+1}(\tilde{D}_{\bst})
    e^{\langle\bsa,D_{\bst}\rangle 
       + \langle\bar{\bsa},D_{\bar{\bst}}\rangle} 
    \tau(s,\bst,\bar{\bst})\cdot\tau(s',\bst,\bar{\bst}) \\
  = \sum_{n=0}^\infty 
    S_n(-2\bar{\bsa})
    S_{n-s'+s-1}(\tilde{D}_{\bar{\bst}})
    e^{\langle\bsa,D_{\bst}\rangle 
       + \langle\bar{\bsa},D_{\bar{\bst}}\rangle} 
    \tau(s-1,\bst,\bar{\bst})\cdot\tau(s'+1,\bst,\bar{\bst}), 
\end{multline}
where $\bsa = (a_1,a_2,\ldots)$ and 
$\bar{\bsa} = (\bar{a}_1,\bar{a}_2,\ldots)$ are auxiliary variables, 
$\langle\bsa,D_{\bst}\rangle$ and 
$\langle\bar{\bsa},D_{\bar{\bst}}\rangle$ are the linear combinations 
\beqnn
  \langle\bsa,D_{\bst}\rangle 
    = \sum_{k=1}^\infty a_kD_k,\quad 
  \langle\bar{\bsa},\bar{D}_{\bar{\bst}}\rangle 
    = \sum_{k=1}^\infty \bar{a}_k\bar{D}_k 
\eeqnn
of $D_k$'s and $\bar{D}_k$'s, 
and $S_n(\tilde{D}_{\bst})$ and $S_n(\tilde{D}_{\bar{\bst}})$ 
are defined by substituting the variables $\bst$ 
of $S_n(\bst)$ for the Hirota bilinear operators 
\beqnn
  \tilde{D}_{\bst} = (D_1,D_2/2,\ldots,D_n/k,\ldots), \quad 
  \tilde{D}_{\bar{\bst}} = (\bar{D}_1,\bar{D}_2/2,\ldots,\bar{D}_k/k,\ldots). 
\eeqnn

\subsection{Orlov-Schulman operators}

Following the idea of Orlov and Schulman \cite{OS86}, 
one can introduce a Toda version of the Orlov-Schulman operator 
of the KP hierarchy.  Actually, we need two Orlov-Schulman 
operators of the form 
\beqnn
\begin{gathered}
  M = \sum_{k=1}^\infty kt_kL^k + s + \sum_{n=1}^\infty v_n L^{-n},\\
  \bar{M} 
    = - \sum_{k=1}^\infty k\bar{t}_k \bar{L}^{-k} + s
      + \sum_{n=1}^\infty \bar{v}_n \bar{L}^n, 
\end{gathered}
\eeqnn
where $v_n$ and $\bar{v}_n$ are new dependent variables.  
These operators are defined in terms of the dressing operators as 
\beq
\begin{gathered}
  M = W\left(s + \sum_{k=1}^\infty kt_ke^{k\rd_s}\right)W^{-1},\\
  \bar{M} 
    = \bar{W}\left(s - \sum_{k=1}^\infty k\bar{t}_ke^{-k\rd_s}\right)\bar{W}^{-1}, 
\end{gathered}
\eeq
and satisfy the Lax equations 
\beq
\begin{gathered}
  \frac{\rd M}{\rd t_n} = [B_n,M], \quad 
  \frac{\rd M}{\rd\bar{t}_n} = [\bar{B}_n,M], \\
  \frac{\rd\bar{M}}{\rd t_n} = [B_n,\bar{M}],\quad
  \frac{\rd\bar{M}}{\rd\bar{t}_n} = [\bar{B}_n,\bar{M}] 
\end{gathered}
\eeq
of the same form as the Lax equations (\ref{Laxeq}) for $L,\bar{L}$. 
Moreover, the twisted canonical commutation relations 
\beq
  [L,M] = L, \quad [\bar{L},\bar{M}] = \bar{L} 
\eeq
are satisfied as a result of the commutation relation (\ref{tCCR}) 
of $e^{\rd_s}$ and $s$. 

These equations form an extended Lax formalism 
of the 2D Toda hierarchy.  One can thereby formulate 
additional symmetries of $W_{1+\infty}$ type \cite{TT93,TT95}. 
These additional symmetries play a central role 
in  the so called ``string equations'' for various special solutions 
\cite{DMP93,HOP94,EK94,Takasaki95,NTT95,Takasaki96,Takasaki12}. 
Moreover, general solutions of the 2D Toda hierarchy, 
too, can be captured by the generalization
\beq
  L = f(\bar{L},\bar{M}),\quad 
  M = g(\bar{L},\bar{M})
  \label{gstreq}
\eeq
of those string equations \cite{TT95,TT12}.

\subsection{Two reductions of 2D Toda hierarchy}

\subsubsection{1D Toda hierarchy}

The 1D Toda hierarchy is a reduction of the 2D Toda hierarchy 
in which all dynamical variables depend on the time variables 
$\bst,\bar{\bst}$ through the difference $\bst - \bar{\bst}$.  
In the Lax formalism, the 1D reduction can be achieved 
by imposing the condition 
\footnote{In the earliest work \cite{UT84}, a condition 
of the form $L + L^{-1} = \bar{L} + \bar{L}^{-1}$ is proposed 
for the 1D reduction.  This condition is related 
to the structure of soliton solutions 
of the Toda lattice \cite{Hirota73,Flaschka74b}.} 
\beq
  L = \bar{L}^{-1}. 
  \label{1D-LLbar}
\eeq
Both sides of this equation become 
a difference operator of the form 
\beq
  \frakL = e^{\rd_s} + b + ce^{-\rd_s}, \quad 
  b = u_1,\quad c = \bar{u}_0, 
  \label{1D-redL}
\eeq
which satisfies the Lax equations 
\beqnn
  \frac{\rd\frakL}{\rd t_k} = [B_k,\frakL], \quad 
  \frac{\rd\frakL}{\rd\bar{t}_k} = [\bar{B}_k,\frakL]. 
\eeqnn 
Since (\ref{1D-LLbar}) implies that $B_k$, $\bar{B}_k$ 
and $\frakL$ are linearly related as 
\beq
  B_k + \bar{B}_k = \frakL^k, 
\eeq
the time evolutions with respect to $\bst$ and $\bar{\bst}$ 
are also linearly related as 
\beqnn
  \frac{\rd\frakL}{\rd t_k} + \frac{\rd\frakL}{\rd\bar{t}_k} 
  = [B_k,\frakL] + [\bar{B}_k,\frakL] 
  = 0. 
\eeqnn
Thus the reduced system has just one set of independent 
Lax equations 
\beqnn
  \frac{\rd\frakL}{\rd t_k} = [B_k,\frakL], \quad 
  B_k = (\frakL^k)_{\ge 0}.  
\eeqnn

\subsubsection{Ablowitz-Ladik hierarchy}

The reduction to the Ablowitz-Ladik hierarchy 
is a kind of ``rational reduction'' \cite{BCR11}. 
This is achieved by assuming that $L$ and $\bar{L}^{-1}$ 
are quotients 
\beq
  L = BC^{-1},\quad \bar{L}^{-1} = CB^{-1} 
\label{AL-LLbar}
\eeq
of two difference operators of the form 
\beqnn
  B = e^{\rd_s} - b,\quad C = 1 - ce^{-\rd_s}. 
\eeqnn
$B^{-1}$ and $C^{-1}$ are understood to be 
difference operators of type $[0,\infty)$ and $(-\infty,0]$. 
More explicitly, 
\beqnn
\begin{gathered}
  B^{-1} = - \sum_{k=0}^\infty (b^{-1}e^{\rd_s})^k b^{-1} 
    = - b(s)^{-1} - \sum_{k=1}^\infty b(s)^{-1}\cdots b(s+k)^{-1}e^{k\rd_s},\\
  C^{-1} = 1 + \sum_{k=1}^\infty (ce^{-\rd_s})^k
    = 1 + \sum_{k=1}^\infty c(s)c(s-1)\cdots c(s-k+1)e^{-k\rd_s}, 
\end{gathered}
\eeqnn
where $b(s)$ and $c(s)$ are abbreviations of 
$c(s,\bst,\bar{\bst})$ and $c(s,\bst,\bar{\bst})$. 
Under this interpretation, $CB^{-1}$ is not the inverse of $BC^{-1}$. 
Thus trivial situation where $L = \bar{L} = e^{\rd_s}$ 
can be avoided.  

The Lax equations (\ref{Laxeq}) of the 2D Toda hierarchy 
can be reduced to (and derived from) the equations 
\beq
\begin{gathered}
  \frac{\rd B}{\rd t_k} 
  = \left((BC^{-1})^k\right)_{\ge 0}B - B\left((C^{-1}B)^k\right)_{\ge 0},\\
  \frac{\rd C}{\rd t_k}
  = \left((BC^{-1})^k\right)_{\ge 0}C - C\left((C^{-1}B)^k\right)_{\ge 0},\\
  \frac{\rd B}{\rd\bar{t}_k}
  = \left((CB^{-1})^k\right)_{<0}B - B\left((B^{-1}C)^k\right)_{<0},\\
  \frac{\rd C}{\rd\bar{t}_k}
  = \left((CB^{-1})^k\right)_{<0}C - C\left((B^{-1}C)^k\right)_{<0}. 
\end{gathered}
\label{BCeq}
\eeq
Note that this is a closed system of evolution equations 
for $B$ and $C$.  This implies that the reduced form (\ref{AL-LLbar}) 
of $L$ and $\bar{L}^{-1}$ is preserved by the time evolutions 
of the 2D Toda hierarchy. 

The reduction condition to the Ablowitz-Ladik hierarchy 
can be reformulated in the alternative form 
\beq
  L = \tilde{C}^{-1}\tilde{B},\quad 
  \bar{L}^{-1} = \tilde{B}^{-1}\tilde{C}, 
  \label{AL-LLbar2}
\eeq
where 
$\tilde{B}$ and $\tilde{C}$ are difference operators 
of the form 
\beqnn
  \tilde{B} = e^{\rd_s} - \tilde{b},\quad 
  \tilde{C} = 1 - \tilde{c}e^{-\rd_s}.  
\eeqnn
Just like $B^{-1}$ and $C^{-1}$ in (\ref{AL-LLbar}), 
$\tilde{B}^{-1}$ and $\tilde{C}^{-1}$ are understood to be 
difference operators of type $[0,\infty)$ and $(-\infty,0]$. 
The Lax equations (\ref{Laxeq}) of the 2D Toda hierarchy
can be reduced to the equations 
\beq
\begin{gathered}
  \frac{\rd\tilde{B}}{\rd t_k} 
  = \left((\tilde{B}\tilde{C}^{-1})^k\right)_{\ge 0}\tilde{B} 
    - \tilde{B}\left(\tilde{C}^{-1}\tilde{B})^k\right)_{\ge 0},\\
  \frac{\rd\tilde{C}}{\rd t_k}
  = \left((\tilde{B}\tilde{C}^{-1})^k\right)_{\ge 0}\tilde{C} 
    - \tilde{C}\left((\tilde{C}^{-1}\tilde{B})^k\right)_{\ge 0},\\
  \frac{\rd\tilde{B}}{\rd\bar{t}_k}
  = \left((\tilde{C}\tilde{B}^{-1})^k\right)_{<0}\tilde{B} 
    - \tilde{B}\left((\tilde{B}^{-1}\tilde{C})^k\right)_{<0},\\
  \frac{\rd\tilde{C}}{\rd\bar{t}_k}
  = \left((\tilde{C}\tilde{B}^{-1})^k\right)_{<0}\tilde{C} 
    - \tilde{C}\left((\tilde{B}^{-1}\tilde{C})^k\right)_{<0} 
\end{gathered}
\eeq
for these operators as well.  

The second reduction condition (\ref{AL-LLbar2}) 
is directly related to an auxiliary linear problem 
of the relativistic Toda hierarchy \cite{Ruijsenaars90}.  
If the Lax operators are factorized in that form, 
the linear equations (\ref{LLbar-Lineq}) for the wave functions 
can be converted to the ``generalized eigenvalue problem'' 
\beq
  \tilde{B}\Psi = z\tilde{C}\Psi, \quad 
  \tilde{B}\bar{\Psi} = z\tilde{C}\bar{\Psi}. 
\eeq
A generalized eigenvalue problem of this form is used 
in Bruschi and Ragnisco's scalar-valued Lax formalism \cite{BR89} 
of the relativistic Toda lattice.  
Moreover, as pointed out by Kharchev et al. \cite{KMZ96}, 
this generalized eigenvalue problem can be derived from 
the traditional $2\times 2$ matrix-valued Lax formalism \cite{AL75} 
of the Ablowitz-Ladik hierarchy.

\subsection{Matrix factorization problem}

General solutions of the 2D Toda hierarchy can be captured 
by a factorization problem \cite{Takasaki84} of the form 
\beq
  \exp\left(\sum_{k=1}^\infty t_k\Lambda^k\right)U
  \exp\left(- \sum_{k=1}^\infty\bar{t}_k\Lambda^{-k}\right) 
  = W^{-1}\bar{W}, 
\label{factor}
\eeq
where $U$ is a given (invertible) constant $\ZZ\times\ZZ$ matrix.  
The problem is to find two $\ZZ\times\ZZ$ matrices 
$W = W(\bst,\bar{\bst})$ and $\bar{W} = \bar{W}(\bst,\bar{\bst})$ 
that are triangular matrices of the form 
\beqnn
  W = 1 + \sum_{n=1}^\infty \diag(w_n(s))\Lambda^{-n},\quad 
  \bar{W} = \sum_{n=0}^\infty\diag(\bar{w}_n(s))\Lambda^n,\quad 
  \bar{w}_0 \not= 0.
\eeqnn
Note that $W$ and $\bar{W}$ amount to the dressing operators 
of the last section by the correspondence 
\beqnn
  A(s,e^{\rd_s}) = \sum_{n\in\ZZ}a_n(s)e^{n\rd_s} 
  \;\longleftrightarrow\;
  A(\Delta,\Lambda) = \sum_{n\in\ZZ}\diag(a_n(s))_{s\in\ZZ}\Lambda^n 
\eeqnn
of difference operators and $\ZZ\times\ZZ$ matrices. 

Since $W$ and $\bar{W}$ are lower and upper triangular matrices, 
the factorization problem (\ref{factor}) 
is an infinite dimensional analogue of the Gauss decomposition 
for finite matrices.  If $W$ and $\bar{W}$ satisfy 
the factorization problem (\ref{factor}), 
one can readily derive the equations 
\beq
\begin{gathered}
  \frac{\rd W}{\rd t_k}W^{-1} + W\Lambda^kW^{-1} 
    = \frac{\rd\bar{W}}{\rd t_k}\bar{W}^{-1},\\
  \frac{\rd W}{\rd\bar{t}_k}W^{-1} 
    = \frac{\rd\bar{W}}{\rd\bar{t}_k}\bar{W}^{-1} 
      + \bar{W}\Lambda^{-k}\bar{W}^{-1}. 
\end{gathered}
\eeq
Splitting these equations to the $(\quad)_{\ge 0}$ 
and $(\quad)_{<0}$ parts, one can see that these equations 
are equivalent to the Sato equations (\ref{Satoeq2}). 
Thus the factorization problem yields a solution 
of the 2D Toda hierarchy. 

In analogy with the procedure of the Gauss decomposition 
for finite matrices, one can express the matrix elements 
of $W$ and $\bar{W}$ as quotients of semi-infinite minors of 
\beq
  U(\bst,\bar{\bst}) 
  = \left(U_{ij}(\bst,\bar{\bst})\right)_{i,j\in\ZZ} 
  = \exp\left(\sum_{k=1}^\infty t_k\Lambda^k\right)U
    \exp\left(- \sum_{k=1}^\infty\bar{t}_k\Lambda^{-k}\right). 
  \label{U(t,tbar)}
\eeq
The common denominator of these quotients 
is a principal minor of $U(\bst,\bar{\bst})$, 
and can be identified with the tau function:
\footnote{This is a place where the aforementioned 
modification of the definition of $\tau(s,\bst,\bar{\bst})$ 
affects the outcome.  In the earlier literature, 
the right hand side of this formula is the minor 
for $i,j < s$ rather than $i,j \leq s$.}
\beq
  \tau(s,\bst,\bar{\bst}) 
  = \det(U_{ij}(\bst,\bar{\bst}))_{i,j\leq s}. 
  \label{tau=det}
\eeq
The determinant expression of the matrix elements 
of $W$ and $\bar{W}$ reproduces 
the generating functional expression 
\beqnn
\begin{gathered}
  1 + \sum_{n=1}^\infty w_nz^{-n} 
  = \frac{\tau(s-1,\bst-[z^{-1}],\bst)}{\tau(s-1,\bst,\bar{\bst})},\\
  \sum_{n=0}^\infty\bar{w}_nz^n 
  = \frac{\tau(s,\bst,\bar{\bst}-[z])}{\tau(s-1,\bst,\bar{\bst})}  
\end{gathered}
\eeqnn
of $w_n$'s and $\bar{w}_n$'s, which implies the relation 
(\ref{Psi-tau-rel}). 

These formal computations can be justified rigorously 
\cite{Takasaki84} in the case where $U$ is given by 
the quotient 
\beq
  U = W_0^{-1}\bar{W}_0
\eeq
of two triangular matrices of the same form as $W$ and $\bar{W}$.  
In this case, $W_0$ and $\bar{W}_0$ can be identified 
with the initial values of $W$ and $\bar{W}$: 
\beq
  W_0 = W|_{\bst=\bar{\bst}=\bszero}, \quad 
  \bar{W}_0 = \bar{W}|_{\bst=\bar{\bst}=\bszero}. 
\eeq
In other words, the factorization problem (\ref{factor}) 
in this setup solves the initial value problem 
of the Sato equations (\ref{Satoeq2}).  

The determinant formula (\ref{tau=det}) has many implications.  
First, this is an analogue of the determinant formula 
of the tau functions of the KP hierarchy. 
Since $U$ is an element of the ``group'' $\GL(\infty)$
\footnote{This notation is used here in a loose sense and 
not intended to denote a true group.}, 
it is $\GL(\infty)$ itself that plays the role 
of the infinite-dimensional Grassmann manifold 
in the case of the KP hierarchy.  More precisely, 
the true phase space lies in the product of 
two flag manifolds in which $W$ and $\bar{W}$ live.  
Second, the generating matrix $U$ is related 
to the generalized string equations (\ref{gstreq}). 
These equations are a consequence of the algebraic relations 
\beq
  \Lambda U = Uf(\Delta,\Lambda),\quad 
  \Delta U = Ug(\Delta,\Lambda) 
\eeq
satisfied by $\Lambda$, $\Delta$ and $U$.  
Third, the determinant formula (\ref{tau=det}) 
can be translated to the language 
of a 2D complex free fermion system.  
Let us turn to this fermionic formalism 
of the 2D Toda hierarchy.

\section{Fermionic formalism} 

\subsection{Complex free fermion system}

Let 
\beqnn
  \psi(z) = \sum_{n\in\ZZ} \psi_nz^{-n-1}, \quad 
  \psi^*(z) = \sum_{n\in\ZZ} \psi^*_nz^{-n} 
\eeqnn
denote the conjugate pair of 2D complex free fermion fields. 
For convenience, we use integers rather than half-integers 
for the labels of Fourier modes $\psi_n,\psi^*_n$. 
The Fourier modes satisfy the anti-commutation relations 
\beqnn
  \psi_m\psi^*_n + \psi^*_n\psi_m = \delta_{m+n,0}, \quad 
  \psi_m\psi_n + \psi_n\psi_m   = 0,\quad 
  \psi^*_m\psi^*_n + \psi^*_n\psi^*_m = 0. 
\eeqnn

$\psi_i$'s and $\psi^*_i$'s are understood to be 
linear operators on the fermionic Fock spaces.  
They act on the Fock space $\calH$ from the left side 
and on its dual space $\calH^*$ from the right side.  
These Fock spaces are decomposed to charge-$s$ sectors 
$\calH_s,\calH^*_s$, $s \in \ZZ$.  
Let $\langle s|$ and $|s\rangle$ denote the ground states 
in $\calH_s$ and $\calH^*_s$:
\footnote{The shift of $s$ in (\ref{Psi-tau-rel}) and (\ref{tau=det}) 
is related to this definition of the ground states.}
\beqnn
  \langle s| = \langle-\infty|\cdots\psi^*_{s-1}\psi^*_{s},\quad 
  |s\rangle = \psi_{-s}\psi_{-s+1}\cdots|-\infty\rangle. 
\eeqnn
Excited states are labelled by partitions $\lambda 
= (\lambda_1,\lambda_2,\cdots,\lambda_n,0,0,\cdots)$, 
$\lambda_1 \geq \lambda_2 \geq \cdots \geq 0$, 
of arbitrary length as 
\beqnn
\begin{gathered}
  \langle\lambda,s| 
 = \langle s|\psi_{-s}\cdots\psi_{n-1-s}
   \psi^*_{\lambda_n-n+1+s}\cdots\psi^*_{\lambda_1+s},\\
  |\lambda,s\rangle 
 = \psi_{-\lambda_1-s}\cdots\psi_{-\lambda_n+n-1-s} 
   \psi^*_{-n+1+s}\cdots\psi^*_{s}|s\rangle. 
\end{gathered}
\eeqnn
$\langle s|$ and $|s\rangle$ are identified with 
$\langle\emptyset,s|$ and $|\emptyset,s\rangle$. 
$|\lambda,s\rangle$ and $\langle\lambda,s|$ represent 
a state in which the semi-infinite subset 
$\{\lambda_i-i+1+s\}_{i=1}^\infty$ of the set $\ZZ$ 
of all energy levels are occupied by particles.   
These vectors form dual bases of $\calH_s$ and $\calH^*_s$: 
\beq
  \langle\lambda,r|\mu,s\rangle = \delta_{\lambda\mu}\delta_{rs}.
\eeq

The normal ordered fermion bilinears 
\beqnn
  {:}\psi_{-i}\psi^*_j{:} 
  = \psi_{-i}\psi^*_j - \langle 0|\psi_{-i}\psi^*_j|0\rangle, 
  \quad i,j \in \ZZ, 
\eeqnn
where
\beqnn
  \langle 0|\psi_{-i}\psi^*_j|0\rangle 
  = \begin{cases}
    1 & \text{if $i = j \leq 0$},\\
    0 & \text{otherwise},
    \end{cases}
\eeqnn
span the one-dimensional central extension $\widehat{\gl}(\infty)$ 
of the Lie algebra $\gl(\infty)$ of $\ZZ\times\ZZ$ matrices 
\cite{Kac-book,MJD-book}.  $\gl(\infty)$ consists of 
infinite matrices $A = (a_{ij})_{i,j\in\ZZ}$ that correspond 
to difference operators of finite type (i.e., of $[M,N]$-type 
for a pair of integers $M,N$ that can depend on $A$). 
For such a matrix $A \in \gl(\infty)$, the fermion bilinear 
\beqnn
  \widehat{A} = \sum_{i,j\in\ZZ}a_{ij}{:}\psi_{-i}\psi^*_j{:} 
\eeqnn
becomes a well-defined linear operator on the Fock space, 
and preserves the charge in the sense that 
\beq
  \langle\lambda,r|\widehat{A}|\mu,s\rangle = 0 
  \quad\text{if $r \not= s$.}
\eeq
The elements of $\widehat{\gl}(\infty)$ satisfy 
the commutation relation
\beq
  [\widehat{A},\widehat{B}] = \widehat{[A,B]} + \gamma(A,B) 
\eeq
with the $c$-number cocycle 
\beq
  \gamma(A,B) = \sum_{i>0,j\leq 0}(a_{ij}b_{ji} - b_{ij}a_{ji}). 
\eeq

\subsection{Vertex operators and Schur functions}

We here introduce the special fermion bilinears 
\beqnn
  J_m = \widehat{\Lambda^m} = \sum_{n\in\ZZ}{:}\psi_{m-n}\psi^*_n{:}, 
  \quad m \in \ZZ, 
\eeqnn
which satisfy the commutation relations 
\beq
  [J_m, J_n] = m\delta_{m+n}
  \label{[J,J]}
\eeq
of the Heisenberg algebra.  These operators are used 
to construct vertex operators.  The matrix elements 
of such a vertex operator with respect to the vectors 
$\langle\lambda,s|$ and $|\mu,s\rangle$ are related 
to the Schur and skew Schur functions.  Actually, 
there are two different types of vertex operators 
that correspond to different formulations of these functions. 

Vertex operators of the first type are given by the product 
\beqnn
  \Gamma_{\pm}(\bsx) = \prod_{i\ge 1}\Gamma_{\pm}(x_i),\quad  
  \bsx = (x_1,x_2,\ldots), 
\eeqnn
of the elementary vertex operators 
\beqnn
  \Gamma_{\pm}(x) 
  = \exp\left(\sum_{k=1}^\infty\frac{x^k}{k}J_{\pm k}\right). 
\eeqnn
The matrix elements of these operators 
are the skew Schur functions $s_{\lambda/\mu}(\bsx)$ 
in the sense of symmetric functions of $\bsx$ \cite{ORV03}: 
\beq
  \langle\lambda,s|\Gamma_{-}(\bsx)|\mu,s\rangle
  = \langle\mu,s|\Gamma_{+}(\bsx)|\lambda,s\rangle
  = s_{\lambda/\mu}(\bsx). 
\eeq
In particular, if $\mu = \emptyset$, the matrix elements 
become the Schur functions $s_\lambda(\bsx)$: 
\beq
  \langle\lambda,s|\Gamma_{-}(\bsx)|s\rangle
  = \langle s|\Gamma_{+}(\bsx)|\lambda,s\rangle
  = s_\lambda(\bsx). 
\eeq

Vertex operators of the second type are defined as 
\beqnn
  \gamma_{\pm}(\bst) = \exp\left(\sum_{k=1}^\infty t_kJ_{\pm k}\right). 
\eeqnn
It is these operators $\gamma_{\pm}(\bst)$ 
that are commonly used in the fermionic formula 
of tau functions of the KP and 2D Toda hierarchies 
\cite{MJD-book,AZ12}. 
The matrix elements of $\gamma_{\pm}(\bst)$ 
are the skew Schur functions $S_{\lambda/\mu}(\bst)$ 
of the $\bst$-variables: 
\beq
  \langle\lambda,s|\gamma_{-}(\bst)|\mu,s\rangle
  = \langle\mu,s|\gamma_{+}(\bst)|\lambda,s\rangle
  = S_{\lambda/\mu}(\bst). 
\eeq
These functions are defined by the determinant formula 
\beq
  S_{\lambda/\mu}(\bst) = \det(S_{\lambda_i-\mu_j-i+j}(\bst))_{i,j=1}^n 
  \label{skewSchur=det}
\eeq
for partitions of the form 
$\lambda = (\lambda_1,\ldots,\lambda_n,0,0,\ldots)$, 
$\mu = (\mu_1,\ldots,\mu_n,0,0,\ldots)$.  
$S_n(\bst)$'s are the polynomials defined by 
the generating function (\ref{S_n}).

$\gamma_{\pm}(\bst)$ can be converted to $\Gamma_{\pm}(\bsx)$ 
by substituting 
\beq
  t_k = \frac{1}{k}\sum_{i\geq 1}x_i^k. 
\eeq
By the same transformation of variables, the polynomials $S_n(\bst)$ 
in $\bst$ turn into the homogeneous symmetric function $h_n(\bsx)$ 
of $\bsx$.  The determinant formula (\ref{skewSchur=det}) 
of $S_{\lambda/\mu}(\bst)$ thereby reproduces the Jacobi-Trudi formula 
of $s_{\lambda/\mu}(\bsx)$.

\subsection{Fermionic formula of tau functions}

In terms of the foregoing vertex operators, 
the fermionic formula of Toda tau functions 
\cite{Takebe91a,Takebe91b,AZ12} reads 
\footnote{A prototype of this formula can be found 
in the work of Jimbo and Miwa \cite{JM83}.}
\beq
  \tau(s,\bst,\bar{\bst}) 
  = \langle s|\gamma_{+}(\bst)g\gamma_{-}(-\bar{\bst})|s\rangle, 
  \label{tau=<..>}
\eeq
where $g$ is an element of the ``group'' $\widehat{\GL}(\infty)$ 
\footnote{This notation, too, is used here in a loose sense 
just like $\GL(\infty)$.}
of Clifford operators (typically, the exponential $e^{\hat{A}}$ 
of a fermion bilinear $\hat{A}$) \cite{Kac-book,MJD-book}. 
Such a Clifford operator induces a linear transformation 
on the linear span of $\psi_i$'s and $\psi^*_i$'s 
by the adjoint action: 
\beq
  g\psi_jg^{-1} = \sum_{i\in\ZZ}\psi_iU_{ij},\quad 
  g\psi^*_jg^{-1} = \sum_{i\in\ZZ}\psi^*_i\tilde{U}_{ij}. 
\eeq
The coefficients $U_{ij}$ and $\tilde{U}_{ij}$ 
satisfy the orthogonality condition
\beq
  \sum_{k\in\ZZ}U_{ik}\tilde{U}_{jk} 
  = \sum_{k\in\ZZ}U_{ki}\tilde{U}_{kj} 
  = \delta_{ij}. 
\eeq
The fermionic formula (\ref{tau=<..>}) corresponds 
to the determinant formula (\ref{tau=det}) 
of the factorization problem (\ref{factor}) 
for the matrix $U = (U_{ij})_{i,j\in\ZZ}$. 

An immediate consequence of (\ref{tau=<..>}) 
is the Schur function expansion 
\beq
  \tau(s,\bst,\bar{\bst}) 
  = \sum_{\lambda,\mu\in\calP}\langle\lambda,s|g|\mu,s\rangle 
    S_\lambda(\bst)S_\mu(-\bar{\bst}), 
\eeq
where $\calP$ denotes the set of all partitions. 
This expansion is obtained by inserting the partition of unity 
\beq
  1 = \sum_{\lambda\in\calP,s\in\ZZ}|\lambda,s\rangle\langle\lambda,s|
  \label{unity}
\eeq
to the two places among $\gamma_{+}(\bst)$, $g$ 
and $\gamma_{-}(-\bar{\bst})$.  
This amounts to applying the Cauchy-Binet formula 
to the determinant formula (\ref{tau=det}).  
The three factors $\langle\lambda,s|g|\mu,s\rangle$, 
$S_\lambda(\bst)$, $S_\mu(-\bar{\bst})$ may be thought of as minors 
of the three matrices on the right side of (\ref{U(t,tbar)}).  
As regards $S_\lambda(\bst)$ and $S_\mu(-\bar{\bst})$, 
this is indeed a consequence of the special case
\beq
  S_\lambda(\bst) = \det(S_{\lambda_i-i+j}(\bst))_{i,j=1}^n
\eeq
of the determinant formula (\ref{skewSchur=det}).

\subsection{Hypergeometric tau functions}

Let us illustrate the fermionic formula (\ref{tau=<..>}) 
in the case of hypergeometric tau functions \cite{OS00,OS01a,OS01b}. 
This is the case where the generating operator $g$ 
of the tau function (\ref{tau=<..>}) corresponds 
to a diagonal matrix in $\GL(\infty)$.  

Let $U = (e^{T_i}\delta_{ij})_{i,j\in\ZZ}$ be such a diagonal matrix.  
The associated generating operator can be expressed as 
\beq
  g = \exp\left(\sum_{n\in\ZZ}T_n{:}\psi_{-n}\psi^*_n{:}\right). 
  \label{HG-g}
\eeq
This operator, too, is diagonal with respect to 
the basis $\{|\lambda,s\rangle\}_{\lambda \in \calP,s \in \ZZ}$, 
of the Fock space. Thus the tau function becomes a single sum 
over all partitions: 
\beq
  \tau(s,\bst,\bar{\bst}) 
  = \sum_{\lambda\in\calP}\langle\lambda,s|g|\lambda,s\rangle
    S_\lambda(\bst)S_\lambda(-\bar{\bst}). 
  \label{HG-tau}
\eeq
The diagonal elements of $g$ takes 
the so called ``contents product'' form: 
\beq
  \langle\lambda,s|g|\lambda,s\rangle 
  = \langle s|g|s\rangle\prod_{(i,j)\in\lambda}r_{j-i+1+s}, 
\eeq
where $(i,j)\in\lambda$ means that $(i,j)$ runs over 
the cells of the Young diagram of shape $\lambda$, 
and $r_n$'s are defined as 
\beqnn
  r_n = e^{T_n - T_{n-1}}
\eeqnn
The $\lambda$-independent factor $\langle s|g|s\rangle$ 
can be expressed as 
\beq
  \langle s|g|s\rangle 
  = \frac{\prod_{i=1}^\infty e^{T_{-i+1+s}}}{\prod_{i=1}^\infty e^{T_{-i+1}}}
  = \begin{cases}
     \prod_{i=1}^s e^{T_{-i+1+s}} & \text{if $s > 0$},\\
     1 & \text{if $s = 0$},\\
     \prod_{i=1}^{-s}e^{-T_{-i+1}} & \text{if $s < 0$}. 
    \end{cases}
  \label{<s|g|s>}
\eeq

These tau functions are called ``hypergeometric'' 
after the work of Orlov and Scherbin \cite{OS00,OS01a,OS01b}, 
because their work aimed at applications 
to multivariate hypergeometric functions.  
Actually, specialization of the parameters $\{T_n\}_{n\in\ZZ}$ 
yields a variety of examples other than hypergeometric functions.  
Earliest examples of these tau functions can be found 
in the studies of random matrix models 
\cite{KMMM93,Orlov02a,Orlov02b,OS05} 
and $c = 1$ string theory \cite{DMP93,NTT95,Takasaki96}. 
Another source of examples is enumerative geometry 
of $\CC\PP^1$ and $\CC^2$ \cite{Okounkov00,LQW03,QW04}. 
Recent researches of this class of tau functions 
are focussed on the double Hurwitz numbers 
\cite{Takasaki12,Alexandrov11,AMMN12,HO15} 
and their variants \cite{GPH14a,GPH14b,Harnad1410,Harnad1504}. 

Let us briefly recall the tau function 
of the double Hurwitz numbers \cite{Okounkov00}. 
The generating operator takes such a form as 
\beq
  g = Q^{L_0}e^{\beta K/2},
\eeq
where $Q$ and $\beta$ are constants, and 
$L_0$ and $K$ are the special fermion bilinears 
\beqnn
\begin{gathered}
  L_0 = \widehat{\Delta} = \sum_{n\in\ZZ}n{:}\psi_{-n}\psi^*_n{:},\\
  K = \widehat{(\Delta-1/2)^2} 
    = \sum_{n\in\ZZ}(n - 1/2)^2{:}\psi_{-n}\psi^*_n{:}. 
\end{gathered}
\eeqnn
The diagonal matrix elements of these fermion bilinears 
can be computed as follows: 
\beq
\begin{gathered}
  \langle\lambda,s|L_0|\lambda,s\rangle =|\lambda| + \frac{s(s+1)}{2},\\
  \langle\lambda,s|K|\lambda,s\rangle 
  = \kappa(\lambda) + 2s|\lambda| + \frac{4s^3-s}{12},
\end{gathered}
\label{<L0>,<K>}
\eeq
where 
\beqnn
\begin{gathered}
  |\lambda| = \sum_{i=1}^\infty\lambda_i,\quad 
  \kappa(\lambda) = \sum_{i=1}^\infty\lambda_i(\lambda_i - 2i + 1). 
\end{gathered}
\eeqnn
This implies that 
\beqnn
\begin{gathered}
  \langle\lambda,s|Q^{L_0}|\lambda,s\rangle = Q^{|\lambda| + s(s+1)/2},\\
  \langle\lambda,s|e^{\beta K/2}|\lambda,s\rangle 
  = e^{\beta(\kappa(\lambda)/2 + s|\lambda| + (4s^3-s)/24)},
\end{gathered}
\eeqnn
Consequently, the tau function has a Schur function expansion 
of the form 
\beq
  \tau(s,\bst,\bar{\bst}) 
  = \sum_{\lambda\in\calP}Q^{|\lambda|+s(s+1)/2}
    e^{\beta(\kappa(\lambda)/2 + s|\lambda| + (4s^3-s)/24)}
    S_\lambda(\bst)S_\lambda(-\bar{\bst}). 
  \label{2Hurwitz-tau}
\eeq
Its specialization 
\beq
  \tau(0,\bst,\bar{\bst}) 
  = \sum_{\lambda\in\calP}Q^{|\lambda|}e^{\beta\kappa(\lambda)/2}
    S_\lambda(\bst)S_\lambda(-\bar{\bst}) 
\eeq
to $s = 0$ is a genuine generating function 
of the double Hurwitz numbers.  

Further specialization to $\bar{\bst} = (-1,0,0,\ldots)$ 
becomes a generating function of the single Hurwitz numbers. 
The special value of the second Schur function at this point 
can be computed by the combinatorial formula \cite{Mac-book}
\beq
  S_\lambda(1,0,0,\ldots) 
  = \frac{\dim\lambda}{|\lambda|!}
  = \prod_{(i,j)\in\lambda}h(i,j)^{-1}, 
  \label{hook-formula}
\eeq
where $h(i,j)$ is the hook length of the cell $(i,j)$ 
in the Young diagram of shape $\lambda$, 
and $\dim\lambda$ is the number of standard tableau therein 
(i.e., the dimension of the associated irreducible 
representation of the symmetric group $S_N$, $N = |\lambda|$).  
The doubly specialized tau function 
\beq
  \tau(0,\bst,-1,0,0,\ldots) 
  = \sum_{\lambda\in\calP}\frac{\dim\lambda}{|\lambda|!}
    Q^{|\lambda|}e^{\beta\kappa(\lambda)/2}S_\lambda(\bst) 
\eeq
reproduces a generating function of the single Hurwitz numbers.  
Note that this is a tau function of the KP hierarchy 
with the fermionic expression 
\beq
  \tau(0,\bst,-1,0,0,\ldots) 
  = \langle 0|\gamma_{+}(\bst)Q^{L_0}e^{\beta K/2}e^{J_{-1}}|0\rangle. 
\eeq

\section{Melting crystal models}

\subsection{Statistical model of random 3D Young diagrams}

The simplest melting crystal model \cite{ORV03} 
has a single parameter $q$ in the range $0 < q < 1$ 
(or just a formal variable). 
The partition function is the sum 
\beq
  Z = \sum_{\pi\in\calP\calP} q^{|\pi|} 
  \label{Z=PPsum}
\eeq
of the Boltzmann weight $q^{|\pi|}$ over the set $\calP\calP$ 
of all plane partitions.  
The plane partition 
\beqnn
  \pi = (\pi_{ij})_{i,j=1}^\infty 
  = \begin{pmatrix}
    \pi_{11} & \pi_{12} & \cdots \\
    \pi_{21} & \pi_{22} & \cdots \\
    \vdots   & \vdots   & \ddots 
    \end{pmatrix}, \quad
  \pi_{i+1,j} \leq \pi_{ij} \geq \pi_{i,j+1}, 
\eeqnn
represent a 3D Young diagram in the first octant of the $xyz$-space. 
$\pi_{ij}$ is the height of the stacks of unit cubes 
on the unit square $[i-1,i]\times[j-1,j]$ of the $xy$-plane.  
$|\pi|$ denotes the volume of the 3D Young diagram, i.e., 
\beqnn
  |\pi| = \sum_{i,j=1}^\infty \pi_{ij}. 
\eeqnn

By the method of diagonal slicing \cite{ORV03}, 
the sum (\ref{Z=PPsum}) over the set of plane partitions 
can be converted to the sum 
\beq
  Z = \sum_{\lambda\in\calP}s_\lambda(q^{-\rho})^2 
  \label{Z=Psum}
\eeq
over the set of ordinary partitions. 
The building block $s_\lambda(q^{-\rho})$ of the Boltzmann weight 
is the special value of the infinite-variate Schur function 
$s_\lambda(\bsx)$ at 
\beqnn
   \bsx = q^{-\rho} = (q^{1/2},q^{3/2},\ldots,q^{i-1/2},\ldots). 
\eeqnn
This is a kind of ``principal specialization'' 
of $s_\lambda(\bsx)$ \cite{Mac-book}, 
and can be computed by the hook-length formula 
\beq
  s_\lambda(q^{-\rho}) 
  = \frac{q^{-\kappa(\lambda)/4}}
    {\prod_{(i,j)\in\lambda}(q^{-h(i,j)/2} - q^{h(i,j)/2})}. 
  \label{q-hook-formula}
\eeq
Note that this formula is a $q$-analogue of 
(\ref{hook-formula}) for $S_\lambda(1,0,0,\ldots)$.  

Let us introduce another parameter $Q$, a discrete variable $s \in \ZZ$ 
and an infinite number of continuous variables $\bst = (t_1,t_2,\ldots)$, 
and deform (\ref{Z=Psum}) as 
\beq
  Z(s,\bst) = \sum_{\lambda\in\calP}s_\lambda(q^{-\rho})^2
            Q^{|\lambda|+s(s+1)/2}e^{\phi(\lambda,s,\bst)}.  
  \label{Z(s)=Psum}
\eeq
$Q^{|\lambda|+s(s+1)/2}$ is the same factor as inserted 
in the tau function (\ref{2Hurwitz-tau}) 
of the double Hurwitz numbers.  
$\phi(\lambda,s,\bst)$ is a linear combination 
\beqnn
  \phi(\lambda,s,\bst) = \sum_{k=1}^\infty t_k\phi_k(\lambda,s) 
\eeqnn
of the external potentials
\beq
  \phi_k(\lambda,s) 
  = \sum_{i=1}^\infty\left(q^{k(\lambda_i-i+1+s)} - q^{k(-i+1+s)}\right) 
    + \frac{1 - q^{ks}}{1 - q^k}q^s, 
  \label{phi_k}
\eeq
and $t_k$'s play the role of coupling constants of these potentials. 
Note that the sum on the right hand side of (\ref{phi_k}) 
is a finite sum, because only a finite number of $\lambda_i$'s 
are non-zero.  (\ref{Z(s)=Psum}) is related to 5D $\calN = 1$ 
supersymmetric $U(1)$ Yang-Mills theory \cite{MNTT04}. 
The external potentials represent the contribution 
of Wilson loops along the fifth dimension therein \cite{NT07}. 

Let us mention that these external potentials are obtained 
from the apparently divergent (as far as $|q| < 1$) expression 
\beq
  \phi_k(\lambda,s) 
  = \sum_{i=1}^\infty q^{k(\lambda_i-i+1+s)} - \sum_{i=1}^\infty q^{k(-i+1)}
  \label{phi_k2}
\eeq
by recombination of terms as 
\beqnn
  \phi_k(\lambda,s) 
  = \sum_{i=1}^\infty\left(q^{k(\lambda_i-i+1+s)} - q^{k(-i+1+s)}\right)
    + \sum_{i=1}^\infty q^{k(-i+1+s)} - \sum_{i=1}^\infty q^{k(-i+1)}. 
\eeqnn
The difference of the last two sums, too, thereby becomes 
a finite sum: 
\beqnn
  \sum_{i=1}^\infty q^{k(-i+1+s)} - \sum_{i=1}^\infty q^{k(-i+1)} \\
  = \left\{\begin{matrix}
     \sum_{i=1}^s q^{k(-i+1+s)} & \text{if $s > 0$}\\
     0 & \text{if $s = 0$}\\
     - \sum_{i=1}^s q^{k(-i+1)} & \text{if $s < 0$}
    \end{matrix}\right\}
  = \frac{1 - q^{ks}}{1 - q^k}q^s. 
\eeqnn
A similar prescription is used in the computation (\ref{<s|g|s>}) 
of the factor $\langle s|g|s\rangle$ in hypergeometric tau functions.  
These computations are related to normal ordering of fermion bilinears. 

It is this deformed partition function $Z(s,\bst)$ 
that is shown to be related to the 1D Toda hierarchy.  
To this end, we use a fermionic expression of $Z(s,\bst)$.  
Before showing this expression, let us present 
another melting crystal model.

\subsection{Modified melting crystal model} 

The second model is obtained by replacing the main part 
of the Boltzmann weight as 
\beqnn
  s_\lambda(q^{-\rho})^2 \;\longrightarrow\; 
  s_\lambda(q^{-\rho})s_{\tp{\lambda}}(q^{-\rho}), 
\eeqnn
where $\tp{\lambda}$ denotes the conjugate (or transposed) partition 
of $\lambda$.  Namely, in place of (\ref{Z=Psum}) or 
its $Q$-deformed version 
\beq
  Z = \sum_{\lambda\in\calP}s_\lambda(q^{-\rho})^2Q^{|\lambda|}, 
  \label{Z(Q)=Psum}
\eeq
we here consider the modified partition function 
\beq
  Z' = \sum_{\lambda\in\calP}s_\lambda(q^{-\rho})s_{\tp{\lambda}}(q^{-\rho})Q^{|\lambda|} 
\eeq
and its deformations by external potentials.  
In view of the relation 
\beqnn
  s_{\tp{\lambda}}(q^{-\rho}) = q^{\kappa(\lambda)/2}s_\lambda(q^{-\rho}) 
\eeqnn
that can be derived from (\ref{q-hook-formula}), 
one can rewrite $Z'$ as 
\beq
  Z' = \sum_{\lambda\in\calP}s_\lambda(q^{-\rho})^2
         q^{\kappa(\lambda)/2}Q^{|\lambda|}.
  \label{Z'(Q)=Psum}
\eeq
These partition functions originate in 
Gromov-Witten/topological string theory 
of special local Calabi-Yau threefolds called 
``local $\CC\PP^1$ geometry'' \cite{BP08,CGMPS06}. 
In particular, $Z'$ is related to the ``resolved conifold'', 
for which Brini pointed out a relation 
to the Ablowitz-Ladik hierarchy \cite{Brini10}.

Let us mention that one can use the homogeneity 
\beqnn
  s_\lambda(Qx_1,Qx_2,\ldots) = Q^{|\lambda|}s_\lambda(x_1,x_2,\ldots) 
\eeqnn
and the Cauchy identities 
\beqnn
\begin{gathered}
  \sum_{\lambda\in\calP}s_\lambda(x_1,x_2,\ldots)s_\lambda(y_1,y_2,\ldots) 
  = \prod_{i,j\geq 1}(1 - x_iy_j)^{-1},\\
  \sum_{\lambda\in\calP}s_\lambda(x_1,x_2,\ldots)s_{\tp{\lambda}}(y_1,y_2,\ldots) 
  = \prod_{i,j\geq 1}(1 + x_iy_j) 
\end{gathered}
\eeqnn
of the Schur functions to convert these partition functions 
to an infinite product form: 
\beq
\begin{gathered}
  Z = \prod_{i,j=1}^\infty (1 - Qq^{i+j-1})^{-1} 
    = \prod_{n=1}^\infty (1 - Qq^n)^{-n}, \\
  Z' = \prod_{i,j=1}^\infty (1 + Qq^{i+j-1})
     = \prod_{n=1}^\infty (1 + Qq^n)^{-n}. 
\end{gathered}
\eeq
These functions are referred to as the ``MacMahon function'' 
in the literature of combinatorics and mathematical physics. 

We deform $Z'$ by two sets of external potentials 
$\phi_{\pm k}(\lambda,s)$, $k = 1,2,\ldots$, 
with coupling constants $\bst = (t_1,t_2,\ldots)$ 
and $\bar{\bst} = (\bar{t}_1,\bar{t}_2,\ldots)$ as
\beq
\begin{gathered}
  Z'(s,\bst,\bar{\bst}) 
  = \sum_{\lambda\in\calP}s_\lambda(q^{-\rho})s_{\tp{\lambda}}(q^{-\rho}) 
    Q^{|\lambda|+s(s+1)/2}e^{\phi(\lambda,s,\bst,\bar{\bst})},\\
  \phi(\lambda,s,\bst,\bar{\bst}) 
  = \sum_{k=1}^\infty t_k\phi_k(\lambda,s) 
    + \sum_{k=1}^\infty\bar{t}_k\phi_{-k}(\lambda,s). 
\end{gathered}
\label{Z'(s)=Psum}
\eeq
$\phi_{-k}(\lambda,s)$'s are defined by the same formula 
as (\ref{phi_k}) with $k$ replaced by $-k$.  
As it turns out, $\bst$ and $\bar{\bst}$ correspond 
to the two sets of time variables of the 2D Toda hierarchy.

\subsection{Fermionic expression of partition functions}

To translate $Z(s,\bst)$ and $Z'(s,\bst,\bar{\bst})$ 
to the language of the complex free fermion system, 
we need some more operators on the Fock space.   

Let us introduce the new fermion bilinears 
\beq
  H_k = \widehat{q^{k\Delta}} 
  = \sum_{n\in\ZZ}q^{kn}{:}\psi_{-n}\psi^*_n{:}, \quad k \in \ZZ. 
  \label{H_k}
\eeq
These operators are diagonal with respect to the basis 
$\{|\lambda,s\rangle\}_{\lambda\in\calP,s\in\ZZ}$, 
and the matrix elements are nothing but 
the external potentials $\phi_k(\lambda,s)$: 
\beq
  \langle\lambda,s|H_k|\lambda,s\rangle = \phi_k(\lambda,s).
\eeq
This explains the origin of the formal expression (\ref{phi_k2}) 
and its interpretation (\ref{phi_k}).  
The exponential factors in (\ref{Z(s)=Psum}) 
and (\ref{Z'(s)=Psum}) can be thereby expressed as 
\beqnn
  e^{\phi(\lambda,s,\bst)} 
  = \langle\lambda,s|e^{H(\bst)}|\lambda,s\rangle,\quad
  e^{\phi(\lambda,s,\bst,\bar{\bst})} 
  = \langle\lambda,s|e^{H(\bst,\bar{\bst})}|\lambda,s\rangle, 
\eeqnn
where 
\beqnn
  H(\bst) = \sum_{k=1}^\infty t_kH_k, \quad 
  H(\bst,\bar{\bst}) 
  = \sum_{k=1}^\infty t_kH_k + \sum_{k=1}^\infty\bar{t}_kH_{-k}. 
\eeqnn
The other building blocks of $Z(s,\bst)$ are similar 
to those of the tau function (\ref{2Hurwitz-tau}) 
of the double Hurwitz numbers: 
\beqnn
\begin{gathered}
  s_\lambda(q^{-\rho}) 
  = \langle s|\Gamma_{+}(q^{-\rho})|\lambda,s\rangle 
  = \langle\lambda,s|\Gamma_{-}(q^{-\rho})|s\rangle,\\
  Q^{|\lambda|+s(s+1)/2} 
  = \langle\lambda,s|Q^{L_0}|\lambda,s\rangle. 
\end{gathered}
\eeqnn
These building bocks are glued together 
by the partition of unity (\ref{unity}) 
to construct the following fermionic formula of $Z(s,\bst)$: 
\beq
  Z(s,\bst) = \langle s|\Gamma_{+}(q^{-\rho})Q^{L_0}
       e^{H(\bst)}\Gamma_{-}(q^{-\rho})|s\rangle
\label{Z(s)=<..>}
\eeq

To derive a similar fermionic formula of $Z'(s,\bst,\bar{\bst})$, 
we use the following variants of $\Gamma_{\pm}(\bsx)$ \cite{YB08}: 
\beqnn
\begin{gathered}
  \Gamma'_{\pm}(\bsx) = \prod_{i\ge 1}\Gamma'_{\pm}(x_i), \quad 
  \bsx = (x_1,x_2,\ldots), \\
  \Gamma'_{\pm}(z) 
  = \exp\left(- \sum_{k=1}^\infty\frac{(-z)^k}{k}J_{\pm k}\right). 
\end{gathered}
\eeqnn
The matrix elements of these modified vertex operators, too, 
are related to the skew Schur functions except that 
they are labelled by conjugate partitions: 
\beq
  \langle\lambda,s|\Gamma'_{-}(\bsx)|\mu,s\rangle 
  = \langle\mu,s|\Gamma'_{+}(\bsx)|\lambda,s\rangle
  = s_{\tp{\lambda}/\tp{\mu}}(\bsx) 
\eeq
Thus the following fermionic formula of $Z'(s,\bst,\bar{\bst})$ 
can be obtained in the same way as the case of $Z(s,\bst)$: 
\beq
  Z'(s,\bst,\bar{\bst}) = \langle s|\Gamma_{+}(q^{-\rho})Q^{L_0}
        e^{H(\bst,\bar{\bst})}\Gamma'_{-}(q^{-\rho})|s\rangle. 
\label{Z'(s)=<..>}
\eeq

These fermionic formulae resemble the fermionic expression 
of the stationary Gromov-Witten invariants of $\CC\PP^1$ 
\cite{OP02a,OP02b} and the instanton partition functions 
of 4D $\calN=2$ supersymmetric gauge theories 
\cite{LMN03,Nekrasov02,NO03,MN06}.  
We use these formulae to show that 
$Z(s,\bst)$ and $Z'(s,\bst,\bar{\bst})$ are related 
to tau functions of the 2D Toda hierarchy.

\section{Integrable structures of melting crystal models}

\subsection{Quantum torus algebra and shift symmetries}

Although the fermionic formulae (\ref{Z(s)=<..>}), 
(\ref{Z'(s)=<..>}) of the partition functions 
of the melting crystal mode resemble 
the fermionic formula (\ref{tau=<..>}) of Toda tau functions, 
they have manifestly different structures.  
In particular, it is $H_k$'s rather than $J_k$'s 
that generate deformations of the partition functions. 
We use special algebraic relations connecting $H_k$'s and $J_k$'s 
to convert the partition functions to Toda tau functions. 
These algebraic relations, referred to as ``shift symmetries'',  
are formulated in the language of a subalgebra 
in $\widehat{\gl}(\infty)$.  

This subalgebra is spanned by the fermion bilinears 
\beqnn
  V^{(k)}_m = q^{-km/2}\widehat{\Lambda^mq^{k\Delta}}
  = q^{-km/2}\sum_{n\in\ZZ}q^{kn}{:}\psi_{m-n}\psi^*_{n}{:}, \quad 
  k,m \in \ZZ. 
\eeqnn
This is substantially the same fermionic realization 
of the quantum torus algebra that are used in the work 
of Okounkov and Pandharipande on $\CC\PP^1$ Gromov-Witten theory 
\cite{OP02a,OP02b}.  $V^{(k)}_m$'s satisfy the commutation relations 
\beq
  [V^{(k)}_m, V^{(l)}_n]
  = (q^{(lm-kn)/2} - q^{(kn-lm)/2})
    \left(V^{(k+l)}_{m+n} - \frac{q^{k+l}}{1-q^{k+l}}\delta_{m+n,0}\right) 
\eeq
for $k$ and $l$ with $k + l \not= 0$ and 
\beq
  [V^{(k)}_m,V^{(-k)}_n] 
  = (q^{-k(m+n)}-q^{k(m+n)})V^{(0)}_{m+n} + m\delta_{m+n,0}. 
\eeq
$H_k$'s and $J_k$'s are particular elements among $V^{(k)}_m$'s: 
\beqnn
  H_k = V^{(k)}_0, \quad J_k = V^{(0)}_k. 
\eeqnn

We have the following three types of shift symmetries 
\cite{NT07,NT08,Takasaki13}: 
\begin{itemize}
\item[(i)]For $k > 0$ and $m \in \ZZ$, 
\begin{multline}
  \Gamma_{-}(q^{-\rho})\Gamma_{+}(q^{-\rho})
   \left(V^{(k)}_m - \frac{q^k}{1-q^k}\delta_{m,0}\right)  \\
  = (-1)^k\left(V^{(k)}_{m+k} - \frac{q^k}{1-q^k}\delta_{m+k,0}\right)
     \Gamma_{-}(q^{-\rho})\Gamma_{+}(q^{-\rho}). 
\label{SSi}
\end{multline}
\item[(ii)] For $k > 0$ and $m \in \ZZ$, 
\begin{multline}
  \Gamma'_{-}(q^{-\rho})\Gamma'_{+}(q^{-\rho})
   \left(V^{(-k)}_m + \frac{1}{1-q^k}\delta_{m,0}\right) \\
  = \left(V^{(-k)}_{m+k} + \frac{1}{1-q^k}\delta_{m+k,0}\right)
     \Gamma'_{-}(q^{-\rho})\Gamma'_{+}(q^{-\rho}). 
\label{SSii}
\end{multline}
\item[(iii)] For $k,m \in \ZZ$, 
\beq
  V^{(k)}_mq^{K/2} = q^{-m/2}q^{K/2}V^{(k+m)}_m. 
\label{SSiii}
\eeq
\end{itemize}
Note that the indices of $V^{(k)}_m$'s are literally 
shifted after exchanging the order of operator product 
with $\Gamma_{-}(q^{-\rho})\Gamma_{+}(q^{-\rho})$, 
$\Gamma'_{-}(q^{-\rho})\Gamma'_{+}(q^{-\rho})$ and $q^{K/2}$. 

In the earlier work \cite{NT07,NT08,Takasaki13}, 
we used the slightly different fermion bilinear 
\beqnn
  W_0 = \sum_{n\in\ZZ} n{:}\psi_{-n}\psi^*_n{:}
\eeqnn
and the algebraic relation 
\beqnn
  V^{(k)}_mq^{W_0/2} = q^{W_0/2}V^{(k+m)}_m 
\eeqnn
in place of $K$ and (\ref{SSiii}).  
This difference does not affect the essential part 
of the whole story.

\subsection{$Z(s,\bst)$ as tau function}

Let us explain how to convert the partition function $Z(s,\bst)$ 
of the first melting crystal model to a tau function 
of the 1D Toda hierarchy with the aid of the foregoing 
shift symmetries \cite{NT07,NT08}. 

The first step is to insert apparently redundant operators 
among $\langle s|$, $|s\rangle$ and the operator product 
in between: 
\begin{multline}
  Z(s,\bst) = q^{-(4s^3-s)/12}
    \langle s|q^{K/2}\Gamma_{-}(q^{-\rho})\Gamma_{+}(q^{-\rho})e^{H(\bst)}\\
  \quad\mbox{}\times 
   Q^{L_0}\Gamma_{-}(q^{-\rho})\Gamma_{+}(q^{-\rho})q^{K/2}|s\rangle. 
  \label{Z(s)-step1}
\end{multline}
This is based on the identities 
\beqnn
\begin{gathered}
  \langle s|q^{K/2} = q^{(4s^3-s)/24}\langle s|,\quad 
  \langle s|\Gamma_{-}(q^{-\rho}) = \langle s|,\\
  q^{K/2}|s\rangle = q^{(4s^3-s)/24}|s\rangle,\quad 
  \Gamma_{+}(q^{-\rho})|s\rangle = |s\rangle 
\end{gathered}
\eeqnn
that can be derived from (\ref{<L0>,<K>}) and the fact 
that $\langle s|J_{-k} = 0$ and $J_k|s\rangle = 0$ for $k > 0$.  
Also note that the order of $Q^{L_0}$ and $e^{H(\bst)}$, 
which are commutative, is reversed. 

The second step is to apply the shift symmetries. 
The first set (\ref{SSi}) of shift symmetries, 
specialized to $m = 0$ and $k > 0$, yields the identity 
\beqnn
  \Gamma_{-}(q^{-\rho})\Gamma_{+}(q^{-\rho}) 
  \left(H_k - \frac{q^k}{1-q^k}\right) 
  = (-1)^kV^{(k)}_k\Gamma_{-}(q^{-\rho})\Gamma_{+}(q^{-\rho}) 
\eeqnn
that connects $V^{(k)}_0 = H_k$ and $V^{(k)}_k$.  
The third set (\ref{SSiii}) of shift symmetries imply 
the relation 
\beqnn
  V^{(k)}_k = q^{k/2}q^{-K/2}J_kq^{K/2} 
\eeqnn
between $V^{(k)}_k$ and $V^{(0)}_k = J_k$. 
Thus $H_k - q^k/(1-q^k)$ and $J_k$ turn out to satisfy 
the intertwining relation 
\beqnn
  q^{K/2}\Gamma_{-}(q^{-\rho})\Gamma_{+}(q^{-\rho})
  \left(H_k - \frac{q^k}{1-q^k}\right) 
  = (-q^{1/2})^kJ_kq^{K/2}\Gamma_{-}(q^{-\rho})\Gamma_{+}(q^{-\rho}). 
\eeqnn
This relation can be exponentiated as 
\begin{multline*}
  q^{K/2}\Gamma_{-}(q^{-\rho})\Gamma_{+}(q^{-\rho})
  \exp\left(\sum_{k=1}^\infty t_k(H_k - \frac{q^k}{1-q^k})\right) \\
  = \exp\left(\sum_{k=1}^\infty (-q^{1/2})^kt_kJ_k\right)
    q^{K/2}\Gamma_{-}(q^{-\rho})\Gamma_{+}(q^{-\rho}). 
\end{multline*}
We can thus rewrite the first half of the operator product 
in (\ref{Z(s)-step1}) as 
\begin{multline}
  q^{K/2}\Gamma_{-}(q^{-\rho})\Gamma_{+}(q^{-\rho})e^{H(\bst)}
   = \exp\left(\sum_{k=1}^\infty \frac{q^kt_k}{1-q^k}\right)\\
  \mbox{}\times 
    \exp\left(\sum_{k=1}^\infty (-q^{1/2})^kt_kJ_k\right)
    q^{K/2}\Gamma_{-}(q^{-\rho})\Gamma_{+}(q^{-\rho}). 
  \label{Z(s)-step2}
\end{multline}

Plugging (\ref{Z(s)-step2}) into (\ref{Z(s)-step1}), 
we obtain the following expression of $Z(s,\bst)$: 
\begin{multline}
  Z(s,\bst) = q^{-(4s^3-s)/12}
    \exp\left(\sum_{k=1}^\infty \frac{q^kt_k}{1-q^k}\right)\\
  \quad \mbox{} \times 
    \langle s|\exp\left(\sum_{k=1}^\infty (-q^{1/2})^kt_kJ_k\right)
    g|s\rangle, 
  \label{Z(s)=tau1}
\end{multline}
where 
\beq
  g = q^{K/2}\Gamma_{-}(q^{-\rho})\Gamma_{+}(q^{-\rho})
      Q^{L_0}\Gamma_{-}(q^{-\rho})\Gamma_{+}(q^{-\rho})q^{K/2}. 
  \label{mcm-g}
\eeq
Let us note that this expression is slightly different 
from the one presented in the previous papers \cite{NT07,NT08}, 
because we use $K$ in place of $W_0$ in (\ref{Z(s)-step1}). 

In much the same way, moving $e^{H(\bst)}$ to the right 
of $\Gamma_{-}(q^{-\rho})\Gamma_{+}(q^{-\rho})q^{K/2}$ 
in (\ref{Z(s)-step1}), we can derive another expression 
of $Z(s,\bst)$: 
\begin{multline}
  Z(s,\bst) = q^{-(4s^3-s)/12}
    \exp\left(\sum_{k=1}^\infty \frac{q^kt_k}{1-q^k}\right)\\
  \quad \mbox{} \times 
    \langle s|g
    \exp\left(\sum_{k=1}^\infty (-q^{1/2})^kt_kJ_{-k}\right)|s\rangle.  
  \label{Z(s)=tau2}
\end{multline}
Actually, as one can show with the aid of the shift symmetries, 
the operator (\ref{mcm-g}) connects $J_k$'s and $J_{-k}$'s as 
\beq
  J_k g = g J_{-k}, \quad k = 1,2,\ldots. 
  \label{mcm-g-symmetry}
\eeq
This explains why $Z(s,\bst)$ has the two apparently different 
expressions (\ref{Z(s)=tau1}) and (\ref{Z(s)=tau2}).

Apart from the prefactors and the rescaling 
$t_k \to (-q^{1/2})^kt_k$ of the time variables, 
the essential part of the right side of (\ref{Z(s)=tau1}) 
and (\ref{Z(s)=tau2}) is the function 
\beq
  \tau(s,\bst) = \langle s|\gamma_{+}(\bst)g|s\rangle 
  = \langle s|g\gamma_{-}(\bst)|s\rangle. 
  \label{1Dtau=<..>}
\eeq
By the symmetry (\ref{mcm-g-symmetry}) of $g$, 
the associated 2D Toda tau function reduces to this function: 
\beq
  \tau(s,\bst,\bar{\bst}) 
  = \langle s|\gamma_{+}(\bst)g\gamma_{-}(-\bar{\bst})|s\rangle 
  = \tau(s,\bst - \bar{\bst}). 
  \label{2D->1Dtau}
\eeq
This means that $\tau(s,\bst)$ is a tau function 
of the 1D Toda hierarchy.   

\begin{remark}
The exponential functions in (\ref{Z(s)=tau1}) and (\ref{Z(s)=tau2}) 
can be absorbed by redefinition of the tau function replacing 
\beqnn
  g \to \tilde{g} 
  = \exp\left(\sum_{k=1}^\infty\frac{(-q^{1/2})^k}{k(1-q^k)}J_{-k}\right)
    g\exp\left(\sum_{k=1}^\infty\frac{(-q^{1/2})^k}{k(1-q^k)}J_k\right).  
\eeqnn
This is a consequence of the identities 
\begin{multline*}
  \exp\left(\sum_{k=1}^\infty \frac{q^kt_k}{1-q^k}\right)
  \exp\left(\sum_{k=1}^\infty(-q^{1/2})^kt_kJ_k\right)\\
  = \exp\left(\sum_{k=1}^\infty(-q^{1/2})^kt_kJ_k\right)
    \exp\left(\sum_{k=1}^\infty\frac{(-q^{1/2})^k}{k(1-q^k)}J_{-k}\right), 
\end{multline*}
\begin{multline*}
  \exp\left(\sum_{k=1}^\infty \frac{q^kt_k}{1-q^k}\right)
  \exp\left(\sum_{k=1}^\infty(-q^{1/2})^kt_kJ_{-k}\right)\\
  = \exp\left(\sum_{k=1}^\infty\frac{(-q^{1/2})^k}{k(1-q^k)}J_k\right) 
    \exp\left(\sum_{k=1}^\infty(-q^{1/2})^kt_kJ_{-k}\right) 
\end{multline*}
that can be deduced from the commutation relations (\ref{[J,J]}) 
of $J_{\pm k}$'s.  Note that the new generating operator $\tilde{g}$, 
too, satisfies the 1D reduction condition 
\beqnn
  J_k\tilde{g} = \tilde{g}J_{-k}, \quad k = 1,2,\ldots. 
\eeqnn
It is also remarkable that the two operators 
in the transformation $g \to \tilde{g}$ 
are related to the vertex operators: 
\beqnn
  \exp\left(\sum_{k=1}^\infty\frac{(-q^{1/2})^k}{k(1-q^k)}J_{\pm k}\right)  
  = \Gamma'_{\pm}(q^{-\rho})^{-1}. 
\eeqnn
\end{remark}

\begin{remark}
There is an another way to avoid the exponential factors 
in (\ref{Z(s)=tau1}) and (\ref{Z(s)=tau2}). These factors 
disappear if the external potentials $\phi_k(\lambda)$ 
are modified as 
\beq
  \phi_k(\lambda,s) 
  = \sum_{i=1}^\infty\left(q^{k(\lambda_i-i+1+s)} - q^{k(-i+1+s)}\right) 
    - \frac{q^{ks}}{1 - q^k}q^s, 
  \label{phi_k-mod}
\eeq
namely, if the constant term $q^k/(1 - q^k)$ is subtracted 
from $\phi_k(\lambda)$.  This amounts to modifying 
the definition (\ref{H_k}) of $H_k$ as 
\beq
  H_k = \widehat{q^{k\Delta}} - \frac{q^k}{1 - q^k}. 
  \label{H_k-mod}
\eeq
The foregoing computations with the aid of the shift symmetries, 
too, can be slightly simplified by this redefinition of $H_k$'s. 
Note that the prefactor $q^{-(4s^3-s)/12}$ cannot be removed 
by this modification. 
\end{remark}

\subsection{$Z'(s,\bst,\bar{\bst})$ as tau function}

The partition function $Z'(s,\bst,\bar{\bst})$ 
of the second melting crystal model can be treated 
in a parallel manner.  Let us show an outline 
of the computations \cite{Takasaki13}. 

The first step is to rewrite the fermionic expression 
(\ref{Z'(s)=<..>}) as follows: 
\begin{multline}
  Z'(s,\bst,\bar{\bst}) 
  = \langle s|q^{K/2}\Gamma_{-}(q^{-\rho})\Gamma_{+}(q^{-\rho})e^{H(\bst)}\\
  \mbox{}\times Q^{L_0}e^{\bar{H}(\bar{\bst})}
    \Gamma'_{-}(q^{-\rho})\Gamma'_{+}(q^{-\rho})q^{-K/2}|s\rangle, 
  \label{Z'(s)-step1}
\end{multline}
where 
\beqnn
  \bar{H}(\bar{\bst}) = \sum_{k=1}^\infty\bar{t}_kJ_{-k}.
\eeqnn
Note that we have split $e^{H(\bst,\bar{\bst})}$ 
into $e^{H(\bst)}$ and $e^{\bar{H}(\bar{\bst})}$, 
and inserted $\Gamma'_{+}(q^{-\rho})q^{-K/2}$ 
in place of $\Gamma_{+}(q^{-\rho})q^{K/2}$ to the right end 
of the operator product.  

The second step is to transfer $e^{H(\bst)}$ and 
$e^{\bar{H}(\bar{\bst})}$ to the left and right ends, 
respectively, with the aid of the shift symmetries.  
Computations for $e^{H(\bst)}$ are exactly the same 
as the case of $Z(s,\bst)$.  To transfer $e^{\bar{H}(\bar{\bst})}$, 
we combine the shift symmetries of the second type (\ref{SSii}) 
and the third type (\ref{SSiii}).  This yields  
the relation 
\beqnn
  \left(H_{-k} + \frac{1}{1-q^k}\right)
  \Gamma'_{-}(q^{-\rho})\Gamma'_{+}(q^{-\rho})q^{-K/2} 
  = \Gamma'_{-}(q^{-\rho})\Gamma'_{+}(q^{-\rho})q^{-K/2}J_{-k} 
\eeqnn
connecting $H_{-k} + 1/(1-q^k)$ and $J_{-k}$. 
Exponentiating this relation, we obtain the following 
counterpart of (\ref{Z'(s)-step2}): 
\begin{multline}
  e^{\bar{H}{(\bar{\bst})}}
  \Gamma'_{-}(q^{-\rho})\Gamma'_{+}(q^{-\rho})q^{-K/2} 
  = \exp\left(- \sum_{k=1}^\infty\frac{\bar{t}_k}{1-q^k}\right)\\
  \mbox{}\times \Gamma'_{-}(q^{-\rho})\Gamma'_{+}(q^{-\rho})q^{-K/2}
    \exp\left(\sum_{k=1}^\infty q^{-k/2}\bar{t}_kJ_{-k}\right). 
  \label{Z'(s)-step2}
\end{multline}

Plugging (\ref{Z(s)-step2}) and (\ref{Z'(s)-step2}) 
into (\ref{Z'(s)-step1}), we can rewrite $Z'(s,\bst,\bar{\bst})$ as 
\begin{multline}
  Z'(s,\bst,\bar{\bst}) 
  = \exp\left(\sum_{k=1}^\infty\frac{q^kt_k-\bar{t}_k}{1-q^k}\right)\\
  \mbox{}\times 
    \langle s|\exp\left(\sum_{k=1}^\infty (-q^{1/2})^kt_kJ_k\right)
    g\exp\left(\sum_{k=1}^\infty q^{-k/2}\bar{t}_kJ_{-k}\right)|s\rangle, 
  \label{Z'(s)=tau}
\end{multline}
where
\beq
  g = q^{K/2}\Gamma_{-}(q^{-\rho})\Gamma_{+}(q^{-\rho})
      Q^{L_0}\Gamma'_{-}(q^{-\rho})\Gamma'_{+}(q^{-\rho})q^{-K/2}. 
  \label{mcm'-g}
\eeq
Thus, apart from the exponential prefactor 
and the rescaling $t_k \to (-q^{1/2})^kt_k$, 
$\bar{t}_k \to q^{-k/2}\bar{t}_k$ of the time variables, 
$Z'(s,\bst,\bar{\bst})$ is a tau function of the 2D Toda hierarchy 
generated by the operator (\ref{mcm'-g}). 

One can find no symmetry like (\ref{mcm-g-symmetry}) 
for the generating operator (\ref{mcm'-g}).  
The associated tau function is a genuine 2D Toda tau function.   
Actually, this special solution of the 2D Toda hierarchy 
falls into the Ablowitz-Ladik hierarchy \cite{Takasaki13}. 

\begin{remark}
The exponential prefactor in (\ref{Z'(s)=tau}) can be absorbed 
by replacing 
\beqnn
  g \to \tilde{g} 
  = \exp\left(\sum_{k=1}^\infty\frac{(-q^{1/2})^k}{k(1-q^k)}J_{-k}\right)
    g\exp\left(- \sum_{k=1}^\infty\frac{q^{k/2}}{k(1-q^k)}J_k\right). 
\eeqnn
Alternatively, one can remove this prefactor 
by subtracting the constant terms $q^{\pm k}/(1 - q^{\pm k})$ 
from the external potentials $\phi_{\pm k}(\lambda)$ as shown 
in (\ref{phi_k-mod}).  The operators $H_{\pm k}$ are 
accordingly modified as shown in (\ref{H_k-mod}). 
\end{remark}

\subsection{Shift symmetries in matrix formalism}

We here turn to a digression on the quantum torus algebra 
and the shift symmetries.  This is not just a digression, 
but closely related to the subsequent consideration 
in the perspective of the Lax formalism.  

The foregoing quantum torus Lie algebra and shift symmetries 
can be translated to the language of infinite matrices 
by the correspondence $A \leftrightarrow \widehat{A}$ 
between $\ZZ\times\ZZ$ matrices and fermion bilinears.  
This matrix formalism enables us to use the associative 
product of matrices as well.  In particular, 
the matrix representation $\bsV^{(k)}_m$ of $V^{(k)}_m$ 
are expressed in term of $\Lambda$ and $\Delta$ as 
\beq
  \bsV^{(k)}_m = q^{-km/2}\Lambda^m q^{k\Delta}. 
  \label{V-matrix}
\eeq
Moreover, the commutation relations 
\beq
  [\bsV^{(k)}_m, \bsV^{(l)}_n] 
  = (q^{(lm-kn)/2} - q^{(kn-lm)/2})\bsV^{(k+l)}_{m+n} 
\eeq
of the centerless quantum torus Lie algebra 
can be derived from the so called quantum torus relation 
\beq
  \Lambda q^\Delta = q q^\Delta\Lambda 
\label{q-torus}
\eeq
satisfied by $\Lambda$ and $q^\Delta$, 
which generate an associative quantum torus algebra.  

Moreover, the vertex operators $\Gamma_{\pm}(q^{-\rho})$ 
and $\Gamma'_{\pm}(q^{-\rho})$ reveals a hidden link 
with the notion of quantum dilogarithmic functions \cite{FV93,FK93} 
through the matrix representation.  Such a Clifford operator, 
too, have the associated matrix representation 
through the exponentiation $e^A \leftrightarrow e^{\hat{A}}$ 
of the Lie algebraic correspondence $A \leftrightarrow \hat{A}$. 
The fundamental vertex operators $\Gamma_{\pm}(x)$ 
and $\Gamma'_{\pm}(x)$ thereby correspond to the following matrices: 
\beq
\begin{gathered}
  \bsGamma_{\pm}(x)
  = \exp\left(\sum_{k=1}^\infty\frac{x^k}{k}\Lambda^{\pm k}\right) 
  = (1 - x\Lambda^{\pm 1})^{-1}, \\
  \bsGamma'_{\pm}(x) 
  = \exp\left(- \sum_{k=1}^\infty\frac{(-x)^k}{k}\Lambda^{\pm}\right)
  = (1 + x\Lambda^{\pm 1}). 
\end{gathered}
\eeq
Consequently, the matrix representation of $\Gamma_{\pm}(q^{-\rho})$ 
and $\Gamma'_{\pm}(q^{-\rho})$ become an infinite product 
of these matrices specialized to $x = q^{i-1/2}$: 
\beq
  \bsGamma_{\pm}(q^{-\rho})
  = \prod_{i=1}^\infty (1 - q^{i-1/2}\Lambda^{\pm 1})^{-1},\quad 
  \bsGamma'_{\pm}(q^{-\rho})
  = \prod_{i=1}^\infty (1 + q^{i-1/2}\Lambda^{\pm 1}). 
\label{GG'-matrix}
\eeq
These infinite products may be thought of 
as matrix-valued quantum dilogarithmic functions 
in the sense of Faddeev et al. 

We thus find the following matrix analogues of the shift symmetries: 
\begin{itemize}
\item[(i)]For $k > 0$ and $m \in \ZZ$, 
\beq
  \bsGamma_{-}(q^{-\rho})\bsGamma_{+}(q^{-\rho})\bsV^{(k)}_m 
  = (-1)^k\bsV^{(k)}_{m+k}\bsGamma_{-}(q^{-\rho})\bsGamma_{+}(q^{-\rho}). 
\eeq
\item[(ii)] For $k > 0$ and $m \in \ZZ$, 
\beq
  \bsGamma'_{-}(q^{-\rho})\bsGamma'_{+}(q^{-\rho})\bsV^{(-k)}_m 
  = \bsV^{(-k)}_{m+k}\bsGamma'_{-}(q^{-\rho})\bsGamma'_{+}(q^{-\rho}). 
\eeq
\item[(iii)] For $k,m \in \ZZ$, 
\beq
  \bsV^{(k)}_mq^{(\Delta-1/2)^2/2} = q^{-m/2}q^{(\Delta-1/2)^2/2}\bsV^{(k+m)}_m. 
  \label{SSiii-matrix}
\eeq
\end{itemize}
These matrix analogues of the shift symmetries can be derived 
from the matrix representation (\ref{V-matrix}), (\ref{GG'-matrix}) 
of $V^{(k)}_m$'s and the vertex operators 
by straightforward computations using the quantum torus relation 
(\ref{q-torus}) \cite{Takasaki13}.

\subsection{Perspectives in Lax formalism}

Let us return to the melting crystal models, 
and consider the associated special solutions 
of the 2D Toda hierarchy in the Lax formalism.  
The goal is to show that the Lax operators $L,\bar{L}$ satisfy 
the reduction conditions (\ref{1D-LLbar}) and (\ref{AL-LLbar2}) 
to the 1D Toda and Ablowitz-Ladik hierarchies \cite{Takasaki13}. 
The reasoning can be outlined as follows. 
\begin{itemize}
\item[1.] It is enough to show that the initial values 
of the Lax operators at $\bst = \bar{\bst} = \bszero$ 
satisfy the reduction condition (\ref{1D-LLbar}) 
and (\ref{AL-LLbar2}), because 
these factorized forms are preserved by the time evolutions 
of the 2D Toda hierarchy.  
\item[2.] One can explicitly solve the factorization problem 
(\ref{factor}) for these cases at the initial time.  
The initial values of the dressing operators are written 
in terms of the matrix representation (\ref{GG'-matrix}) 
of the vertex operators and some other simple matrices.  
\item[3.] The initial values of the Lax operators can be computed 
with the aid of these matrices, and turn out to take 
the forms as shown in (\ref{1D-LLbar}) and (\ref{AL-LLbar2}). 
\end{itemize}

\subsubsection{First melting crystal model}

The generating operator (\ref{mcm-g}) in this case 
corresponds to a matrix of the form 
\beq
  U = q^{(\Delta-1/2)^2/2}\bsGamma_{-}(q^{-\rho})\bsGamma_{+}(q^{-\rho})
      Q^\Delta\bsGamma_{-}(q^{-\rho})\bsGamma_{+}(q^{-\rho})q^{(\Delta-1/2)^2/2}. 
  \label{mcm-U}
\eeq
One can use the identities 
\beq
  Q^\Delta\Lambda^n Q^{-\Delta} = Q^{-n}\Lambda^n,\quad 
  Q^{-\Delta}\Lambda^n Q^\Delta = Q^n\Lambda^n 
  \label{Q^D-scaling}
\eeq
to rewrite the triple product in the middle as 
\beqnn
  U = q^{(\Delta-1/2)^2/2}\bsGamma_{-}(q^{-\rho})\bsGamma_{-}(Qq^{-\rho})
      Q^\Delta\bsGamma_{+}(Qq^{-\rho})\bsGamma_{+}(q^{-\rho})q^{(\Delta-1/2)^2/2}. 
\eeqnn
This matrix is already factorized to a product 
of lower and upper triangular matrices as 
\beqnn
  U = W_0^{-1}\bar{W}_0, 
\eeqnn
where 
\beq
\begin{gathered}
  W_0 = q^{(\Delta-1/2)^2/2}\bsGamma_{-}(Qq^{-\rho})^{-1}
      \bsGamma_{-}(q^{-\rho})^{-1}q^{-(\Delta-1/2)^2/2},\\
  \bar{W}_0 = q^{(\Delta-1/2)^2/2}Q^\Delta\bsGamma_{+}(Qq^{-\rho})
            \bsGamma_{+}(q^{-\rho})q^{(\Delta-1/2)^2/2}. 
\end{gathered}
\label{mcm-WWbar0}
\eeq
This means that $W_0$ and $\bar{W}_0$ are the initial values 
$W|_{\bst=\bar{\bst}=\bszero}$, $\bar{W}|_{\bst=\bar{\bst}=\bszero}$ 
of the dressing operators determined 
by the generating matrix (\ref{mcm-U}).  

One can compute the initial values 
\beqnn
  L_0 = L|_{\bst=\bar{\bst}=\bszero} = W_0\Lambda W_0^{-1}, \quad 
  \bar{L}_0^{-1} = \bar{L}|_{\bst=\bar{\bst}=\bszero}^{-1} 
     = \bar{W}_0\Lambda^{-1}\bar{W}_0^{-1}
\eeqnn
of the Lax operators from these explicit forms 
of $W_0$ and $\bar{W}_0$ as follows.  

The first step for computing $L_0$ is to uses the identity 
\beqnn
  q^{-(\Delta-1/2)^2/2}\Lambda q^{(\Delta/1/2)^2/2} = q^\Delta\Lambda 
\eeqnn
that is a consequence of (\ref{SSiii-matrix}).  
By this identity and the expression (\ref{mcm-WWbar0}) 
of $W_0$, one can rewrite $L_0$ as 
\beqnn
  L_0 = q^{(\Delta-1/2)^2/2}\bsGamma_{-}(Qq^{-\rho})^{-1}\bsGamma_{-}(q^{-\rho})^{-1}
        q^\Delta\Lambda\bsGamma_{-}(q^{-\rho})\bsGamma_{-}(Qq^{-\rho})
        q^{-(\Delta-1/2)^2/2}. 
\eeqnn
Since $\bsGamma_{-}(q^{-\rho})$ and $\bsGamma_{-}(Qq^{-\rho})$ 
are matrices of the form 
\beqnn
  \bsGamma_{-}(q^{-\rho})
  = \prod_{i=1}^\infty (1 - q^{i-1/2}\Lambda^{-1})^{-1},\quad
  \bsGamma_{-}(Qq^{-\rho})
  = \prod_{i=1}^\infty (1 - Qq^{i-1/2}\Lambda^{-1})^{-1}, 
\eeqnn
the matrix $\Lambda$ in front of these two matrices 
can be moved to the right side as 
\beqnn
  \Lambda\bsGamma_{-}(q^{-\rho})\bsGamma_{-}(Qq^{-\rho})
  = \bsGamma_{-}(q^{-\rho})\bsGamma_{-}(Qq^{-\rho})\Lambda. 
\eeqnn

One can further use the identity 
\beqnn
  q^\Delta \Lambda^{-1}q^{-\Delta} = q\Lambda^{-1} 
\eeqnn
to transfer the remaining $q^\Delta$ to the right as 
\beqnn
\begin{aligned}
  q^\Delta\bsGamma_{-}(q^{-\rho})\bsGamma_{-}(Qq^{-\rho})
  &= q^\Delta\prod_{i=1}^\infty(1 - q^{i-1/2}\Lambda^{-1})^{-1}
     \prod_{i=1}^\infty(1 - Qq^{i-1/2}\Lambda^{-1})^{-1} \\
  &= \prod_{i=1}^\infty(1 - q^{i+1/2}\Lambda^{-1})^{-1}
     \prod_{i=1}^\infty(1 - Qq^{i+1/2}\Lambda^{-1})^{-1}\cdot q^\Delta \\
  &= \bsGamma_{-}(q^{-\rho})\bsGamma_{-}(Qq^{-\rho})
     (1 - Qq^{1/2}\Lambda^{-1})(1 - q^{1/2}\Lambda^{-1})q^\Delta. 
\end{aligned}
\eeqnn
The outcome reads 
\beqnn
  L_0 = q^{(\Delta-1/2)^2/2}(1 - Qq^{1/2}\Lambda^{-1})(1 - q^{1/2}\Lambda^{-1})
        q^\Delta\Lambda q^{-(\Delta-1/2)^2/2}. 
\eeqnn

Lastly, by the identities 
\beqnn
\begin{gathered}
  q^{(\Delta-1/2)^2/2}\Lambda q^{-(\Delta-1/2)^2/2} = q^{-\Delta}\Lambda,\\
  q^{(\Delta-1/2)^2/2}\Lambda^{-1}q^{-(\Delta-1/2)^2/2} 
   = \Lambda^{-1}q^\Delta = q^{-1}q^\Delta\Lambda^{-1} 
\end{gathered}
\eeqnn
one can rewrite the last expression of $L_0$ as 
\beq
\begin{aligned}
  L_0 &= (1 - Qq^{-1/2}q^\Delta\Lambda^{-1})
         (1 - q^{-1/2}q^\Delta\Lambda^{-1})\Lambda\\
  &= \Lambda - (Q+1)q^{-1/2}q^\Delta + Qq^{-2}q^{2\Delta}\Lambda^{-1}. 
\end{aligned} 
\label{mcm-L0}
\eeq

One can compute $\bar{L}_0^{-1}$ in much the same way, 
and confirm that it coincides with the expression (\ref{mcm-L0}) 
of $L_0$.  This implies that the reduction condition (\ref{1D-LLbar}) 
to the 1D Toda hierarchy is indeed satisfied.

\subsubsection{Second melting crystal model}

The generating operator (\ref{mcm'-g}) in this case 
corresponds to the matrix 
\beq
  U = q^{(\Delta-1/2)^2/2}\bsGamma_{-}(q^{-\rho})\bsGamma_{+}(q^{-\rho})
      Q^\Delta\bsGamma'_{-}(q^{-\rho})\bsGamma'_{+}(q^{-\rho})q^{-(\Delta-1/2)^2/2}.
  \label{mcm'-U} 
\eeq
This matrix can be factorized as 
\beqnn
  U = W_0^{-1}\bar{W}_0 
\eeqnn
with 
\beq
\begin{gathered}
  W_0 = q^{(\Delta-1/2)^2/2}\bsGamma'_{-}(Qq^{-\rho})^{-1}
      \bsGamma_{-}(q^{-\rho})^{-1}q^{-(\Delta-1/2)^2/2},\\
  \bar{W}_0 = q^{(\Delta-1/2)^2/2}Q^\Delta\bsGamma_{+}(Qq^{-\rho})
            \bsGamma'_{+}(q^{-\rho})q^{-(\Delta-1/2)^2/2}. 
\end{gathered}
\label{mcm'-WWbar0}
\eeq

One can compute $L_0$ in much the same way as the previous case, 
starting from the expression 
\beqnn
  L_0 = q^{(\Delta-1/2)^2/2}\bsGamma'_{-}(Qq^{-\rho})^{-1}\bsGamma_{-}(q^{-\rho})^{-1}
        q^\Delta\Lambda\bsGamma_{-}(q^{-\rho})\bsGamma'_{-}(Qq^{-\rho})
        q^{-(\Delta-1/2)^2/2}. 
\eeqnn
This expression contains 
\beqnn
  \bsGamma'_{-}(Qq^{-\rho})
  = \prod_{i=1}^\infty (1 + Qq^{i-1/2}\Lambda^{-1})
\eeqnn
in place of $\bsGamma_{-}(Qq^{-\rho})$.  
Consequently, the foregoing transfer procedure of $q^\Delta$ 
is modified as 
\beqnn
\begin{aligned}
  q^\Delta\bsGamma_{-}(q^{-\rho})\bsGamma'_{-}(Qq^{-\rho})
  &= q^\Delta\prod_{i=1}^\infty(1 - q^{i-1/2}\Lambda^{-1})^{-1}
     \prod_{i=1}^\infty(1 + Qq^{i-1/2}\Lambda^{-1}) \\
  &= \prod_{i=1}^\infty(1 - q^{i+1/2}\Lambda^{-1})^{-1}
     \prod_{i=1}^\infty(1 + Qq^{i+1/2}\Lambda^{-1})\cdot q^\Delta \\
  &= \bsGamma_{-}(q^{-\rho})\bsGamma'_{-}(Qq^{-\rho})
     (1 + Qq^{1/2}\Lambda^{-1})^{-1}(1 - q^{1/2}\Lambda^{-1})q^\Delta. 
\end{aligned}
\eeqnn
The final expression of $L_0$ takes the quotient form 
\beq
  L_0 = (1 + Qq^{-1/2}q^\Delta\Lambda^{-1})^{-1}
        (1 - q^{-1/2}q^\Delta\Lambda^{-1})\Lambda. 
  \label{mcm'-L0}
\eeq

One can compute $\bar{L}_0^{-1}$ in much the same 
(but slightly more complicated) way starting.  
(\ref{mcm'-WWbar0}) and the identity 
\beqnn
  q^{-(\Delta-1/2)^2/2}\Lambda^{-1}q^{(\Delta-1/2)^2/2} 
  = \Lambda^{-1}q^{-\Delta} 
\eeqnn
imply that $\bar{L}_0^{-1}$ can be expressed as 
\begin{multline*}
  \bar{L}_0^{-1} 
  = q^{(\Delta-1/2)^2/2}Q^\Delta\bsGamma_{+}(Qq^{-\rho})\bsGamma'_{+}(q^{-\rho})\\
  \mbox{}\times 
    \Lambda^{-1}q^{-\Delta}\bsGamma'_{+}(q^{-\rho})^{-1}\bsGamma_{+}(Qq^{-\rho})^{-1}
    Q^{-\Delta}q^{-(\Delta-1/2)^2/2}. 
\end{multline*}
The outcome of somewhat lengthy computations reads 
\beq
  \bar{L}_0^{-1} = (1 - q^{1/2}q^{-\Delta}\Lambda)^{-1}
    (1 + Q^{-1}q^{1/2}q^{-\Delta}\Lambda)Q\Lambda^{-1}. 
  \label{mcm'-Lbar0}
\eeq
It is easy to see that (\ref{mcm'-L0}) and 
(\ref{mcm'-Lbar0}) can be rewritten as 
\beqnn
  L_0 = \tilde{C}_0^{-1}\tilde{B}_0, \quad 
  \bar{L}_0^{-1} = - \tilde{B}_0^{-1}\tilde{C}_0, 
\eeqnn
where 
\beq
  \tilde{B}_0 = \Lambda - q^{-1/2}q^\Delta,\quad 
  \tilde{C}_0 = 1 + Qq^{-1/2}q^\Delta\Lambda^{-1}. 
\eeq
This coincides with the reduced form of (\ref{AL-LLbar2}) 
except for the negative sign in the expression of $\bar{L}_0^{-1}$.  
The negative sign is harmless, because it can be absorbed 
by the time reversal $\bar{\bst} \to -\bar{\bst}$.  
Actually, one can express $L_0$ and $\bar{L}_0$ 
in the form of (\ref{AL-LLbar}) as well 
(again with an extra negative sign) \cite{Takasaki13}. 
Anyway, the reduction condition to the Ablowiz-Ladik hierarchy 
is satisfied in this case.

\section{Conclusion}

It is remarkable that the two melting crystal models 
repeat the same pattern of integrable structures 
as the Hermitian and unitary matrix models.  
A major difference is the fact that the partition functions 
of the matrix models are $s \times s$ determinants 
(hence the lattice coordinate $s$ therein take values 
in positive integers), whereas there is no such expression 
of the partition functions of the melting crystal models 
as determinants of finite size. The discrete variable $s$ 
of the melting crystal models enters the Boltzmann weights 
as a parameter.  This is a main reason why we need 
an entirely different method to identify 
the underlying integrable structures.  

On the other hand, the undeformed partition functions 
(\ref{Z(Q)=Psum}) and (\ref{Z'(Q)=Psum}) 
of the two melting crystal models differs 
in just the single factor $q^{\kappa(\lambda)/2}$. 
It is somewhat surprising that this tiniest modification 
leads to a drastic change in the underlying integrable structure.  
Of course this is rather natural from a geometric point of view, 
because the associated Calabi-Yau threefolds are different. 

The shift symmetries of the quantum torus algebra lie 
in the heart of our method.  These algebraic relations 
are used to transform the ``diagonal'' Hamiltonians $H_k = V^{(k)}_0$ 
to the ``non-diagonal'' generators $J_m$ of time evolutions 
of the 2D Toda hierarchy.  Let us mention two other approaches 
to this kind of unconventional time evolutions 
(see also Section 3.5 of the review of Alexandrov 
and Zabrodin \cite{AZ12}). 

The first one is Orlov's approach \cite{Orlov03} 
to a class of KP tau functions obtained 
from the hypergeometric functions (\ref{HG-tau}) 
by specializing the second set $\bar{\bst}$ 
of time variables to a particular point. 
The special value of $S_\lambda(-\bar{\bst})$ 
at that point $\bar{\bst} = - \bsa$ becomes 
a determinant of the Cauchy type. 
The Schur function expansion of $\tau(s,\bst,-\bsa)$ 
can be thereby reorganized to an ``$\infty$-soliton solution'' 
of the KP hierarchy in which the parameters 
$\bsT = (T_1,T_2,\ldots)$ of the generating operator (\ref{HG-g}) 
play the role of time variables.  

The second approach is developed by Bettelheim et al. \cite{BAW06} 
in their research of a complex fermion system on the real line. 
Time evolutions of this system are generated by 
diagonal Hamiltonians similar to our $H_k$'s except that 
the coefficients $q^{kn}$ of ${:}\psi^*_{-n}\psi_n{:}$ 
are replaced by $n^k$.  Bettelheim et al. considered 
an analogue of KP and Toda tau functions in which 
$J_k$'s and the ground states $\langle s|$ and $|s\rangle$ 
are replaced by $H_k$'s and what they call ``boundary states'', 
$\langle B_s|$ and $|B_s\rangle$. 
These boundary states are generated from the vacuum states 
$\langle 0|$ and $|0\rangle$ by ``boundary operators'' $B_s$. 
The modified ``tau functions'' are shown to satisfy 
the bilinear equations of the KP and Toda hierarchies.  
Unfortunately, it is difficult to compare the results 
of Bettelheim et al. with ours literally, 
because the setup of the fermion system is different.  
Our complex free fermions live on a circle $|z| = R$ 
of the $z$-plane rather than the real axis. 
Nevertheless, it is obvious that the boundary operators $B_s$ 
play the same role as $\Gamma_{-}(q^{-\rho})\Gamma_{+}(q^{-\rho})$ 
in our approach.  

We believe that the shift symmetries will be useful 
beyond the scope of the melting crystal models.  
The results reviewed in this paper should be 
just a small piece of possible applications.  In fact, 
we recently applied the shift symmetries to computations 
of topological string theory in a special case \cite{NT15}.  
We are currently trying to find how the algebraic relations 
(\ref{SSi}) and (\ref{SSii}) are altered outside the range $k > 0$. 
Hopefully, the shift symmetries thus extended will become 
a new tool of computations for various purposes. 

It is also true that the shift symmetries 
are a very special property of the vertex operators 
$\Gamma_{\pm}(q^{-\rho})$ and $\Gamma'_{\pm}(q^{-\rho})$.  
Until now, we have been unable to find a similar tool 
for the 4D version \cite{LMN03,Nekrasov02,NO03,MN06} 
of $Z(s,\bst)$ and $Z'(s,\bst,\bar{\bst})$.  
The aforementioned boundary operators of Bettelheim et al.  
might be a clue to this problem.  It seems more likely 
that another clue is hidden in the fermionic formalism 
of $\CC\PP^1$ Gromov-Witten theory developed 
by Okounkov and Pandharipande \cite{OP02a,OP02b}.

\subsection*{Acknowledgements}

I would like to thank Takashi Takebe and Toshio Nakatsu 
for longstanding collaboration.  I am also grateful 
to Mark Adler, Pierre van Moerbeke, John Harnad 
and Sasha Orlov for having constant interests 
in the Toda hierarchy.  Last but not least, 
I am indebted to Kimio Ueno for support in the earliest stage 
of the studies on the Toda hierarchies.  
This work is partly supported by the JSPS Kakenhi Grant 
No. 25400111 and No. 15K04912.

\end{document}